\documentclass[conference,compsoc]{IEEEtran}

\ifCLASSOPTIONcompsoc
  \usepackage[nocompress]{cite}
\else
  \usepackage{cite}
\fi

\ifCLASSINFOpdf
  \usepackage[pdftex]{graphicx}
  \graphicspath{{./}}
  \usepackage[T1]{fontenc}
  \usepackage{roboto}
  \usepackage{tikz}
  \usetikzlibrary{
      arrows.meta,
      calc,
      shapes.geometric,
      positioning,
      decorations.pathreplacing
  }
\else
\fi

\usepackage{listings}
\usepackage{listingsutf8}
\lstdefinelanguage{Markdown}{
  morecomment=[l][\color{blue!65!black}\bfseries]{\#},
  morecomment=[s][\color{black!70}]{```}{```},
  morestring=[b][\color{magenta!70!black}]{`},
  moredelim=[s][\bfseries\color{blue!80!black}]{**}{**},
  moredelim=[s][\itshape\color{blue!80!black}]{*}{*}
}
\usepackage{colortbl}
\definecolor{deltapos}{HTML}{1B7837}
\definecolor{deltaneg}{HTML}{B2182B}
\usepackage{array}
\usepackage{booktabs}
\usepackage{multirow}
\usepackage{tabularx}
\usepackage{tablefootnote}
\usepackage{caption}
\usepackage{float}
\newfloat{listing}{tbp}{lol}
\floatname{listing}{Listing}
\usepackage[skins]{tcolorbox}
\lstdefinestyle{ccode}{language=C,basicstyle=\ttfamily\footnotesize,backgroundcolor=\color{gray!10},frame=tb,framesep=2pt,numbers=left,numbersep=3pt,numberstyle=\tiny,xleftmargin=0pt,framexleftmargin=0pt,breaklines=true,tabsize=2,keywordstyle=\color{blue!65!black}\bfseries,commentstyle=\color{green!45!black}\itshape,stringstyle=\color{magenta!70!black}}
\newcommand{\designcircnum}[1]{%
  \tikz[baseline=(n.base)]\node[
    circle,
    draw=black,
    text=black,
    fill=white,
    minimum size=1.2em,
    inner sep=1pt,
    font=\sffamily\bfseries\footnotesize
  ] (n) {#1};%
}

\usepackage{xspace}
\newcommand{\sys}{SCARVA\xspace}

\usepackage[hidelinks]{hyperref}
\makeatletter
\renewcommand{\sectionautorefname}{\S\@gobble}
\renewcommand{\subsectionautorefname}{\S\@gobble}
\renewcommand{\subsubsectionautorefname}{\S\@gobble}

\makeatother
\let\designcircnumorig\designcircnum
\renewcommand{\designcircnum}[1]{\texorpdfstring{\designcircnumorig{#1}}{}}

\newcommand{\eg}{\emph{e.\,g.},\xspace}
\newcommand{\ie}{\emph{i.\,e.},\xspace}
\newcommand{\etal}{\emph{et al.}\xspace}

\newcommand{\analyst}{\textsc{analyst}\xspace}
\newcommand{\judge}{\textsc{judge}\xspace}

\newcommand{\gemma}{Gemma 4\xspace}
\newcommand{\muse}{Muse Glimmer\xspace}
\newcommand{\qwen}{Qwen 3.8\xspace}
\newcommand{\luna}{GPT-5.6 Luna\xspace}
\newcommand{\terra}{GPT-5.6 Terra\xspace}
\newcommand{\sol}{GPT-5.6 Sol\xspace}
\newcommand{\opus}{Opus 5.5\xspace}
\newcommand{\sonnet}{Sonnet 5.5\xspace}
\newcommand{\haiku}{Haiku 5.5\xspace}

\newcommand{\finding}[1]{%
  \begin{tcolorbox}[
      enhanced,
      colback=gray!6, colframe=black!65,
      colbacktitle=blue!15, coltitle=black,
      title={Takeaway},
      fonttitle=\sffamily\bfseries\fontsize{8}{8}\selectfont,
      boxrule=0.5pt, arc=2mm, outer arc=2mm,
      toptitle=0.1mm, bottomtitle=0.1mm,
      left=3pt, right=3pt, top=2pt, bottom=2pt,
      before skip=4pt, after skip=4pt,
      fontupper=\small]
    #1
  \end{tcolorbox}}

\newcommand{\dg}[1]{\mbox{D-\ifcase#1\or TP\or FP\or TN\fi}}

\newif\ifanon
\anonfalse

\begin{document}

\title{\sys: Interpretable Agentic Reasoning in Software Supply-Chain Analysis}

\author{
    \IEEEauthorblockN{
        Tim Barsch\IEEEauthorrefmark{1}, 
        Bruno Nogueira\IEEEauthorrefmark{1}, 
        Tobias Wienand\IEEEauthorrefmark{1},
        Jan Fopma\IEEEauthorrefmark{1},
        Flavio Toffalini\IEEEauthorrefmark{1}
    }
\IEEEauthorblockA{\IEEEauthorrefmark{1}
    Ruhr University Bochum}
}

\maketitle

\begin{abstract}
Complex applications may inherit security vulnerabilities from their
dependencies, though a vulnerable dependency does not necessarily endanger the
application.
Existing tools flag dependencies associated with public vulnerabilities, and
more advanced ones analyze the call graph to assess whether the vulnerable code
is reachable.
However, these approaches are limited to specific programming languages and
ecosystems.
LLM agents are promising candidates for this task, yet they hallucinate, take
shortcuts, and reward hack.
As these behaviors are opaque, the agents' conclusions may not be trusted.

We present \sys, a framework designed to produce interpretable reports.
\sys combines advanced agent reasoning with an evidence-collection methodology
that assesses the agent's analysis and improves its trustworthiness.
Specifically, the \analyst performs a software security assessment while \sys
records every tool call, file access, and web request.
Then, one or more independent \judge agents inspect the \analyst's result and
produce an evidence-backed report.
We evaluate \sys on 54 manually curated cases, balanced across exploitable,
non-exploitable, and vulnerability-free projects, using both open-weight and
frontier models.
The best \analyst model, Qwen 3.8, reaches 82\% balanced accuracy and surpasses
the evaluated proprietary models.
Human annotation validates the \judge findings: Claude Opus 5.5 performs best,
agreeing with annotators in 94\% of claim checks.
Finally, we discuss the most common failure modes of agents on supply-chain
vulnerability assessment.
\end{abstract}


\IEEEpeerreviewmaketitle

\section{Introduction}

Modern software projects rely on a complex web of third-party libraries to
implement specific tasks, such as cryptographic algorithms, network
communication, or image processing.
The set of libraries used in an application defines its software supply chain.
Software dependencies allow developers to reuse existing solutions, and thus
save time.
However, dependencies may silently hide security vulnerabilities that the
application inherits~\cite{understanding-software-supplychain}.
Therefore, tracing vulnerabilities and attributing them to a specific
third-party library version is crucial for assessing the security of an
application using that library.
To track publicly disclosed vulnerabilities, the security community maintains
the Common Vulnerabilities and Exposures (CVE) Program, in which every new issue
receives a unique CVE identifier.
The corresponding CVE record documents the affected library versions, the first
patched version, and a description of the issue.
Whenever a new CVE is published, developers must answer a pressing question:
\emph{Does this CVE affect my application?}

The community addresses this problem with software supply-chain analysis,
implemented by Software Bill of Materials (SBOM) tools~\cite{syft, trivy,
dependencytrack}.
SBOM tools enumerate the dependencies of a project and match their versions
against public CVEs~\cite{grype, osvscanner}.
However, a library version that matches a CVE does not imply that the CVE is a
threat to the application, because exploitation also requires an
attacker-controlled input that reaches the vulnerable code and the runtime
conditions triggering the flaw.
The academic community therefore explores more sophisticated approaches that
reason about the code base.
For instance, graph-based approaches assess reachability, but they mostly target
ecosystems with centralized package management, such as Maven, npm, or PyPI,
where dependency and call graphs are readily available~\cite{vuln4real,
vultracer}.
C and C++ projects, where libraries are vendored, built conditionally, and
selected at runtime by configuration, remain largely out of reach for these
approaches, and their exploit-specific data flow is rarely analyzed.
In practice, the verdict is delegated to a human analyst, who reads
both code bases and builds a convincing argument about whether a viable path to
the vulnerable code exists.

Recent work shows that LLM agents achieve strong results on code comprehension
and vulnerability analysis~\cite{do-llm-their-job, code-analysis-llm}, which
suggests that this manual effort can be offloaded to an agent.
Such an agent inspects the application and the library, reasons about the attack
surface, and concludes with one of two answers: the application is
\emph{affected}, together with a proof of concept that triggers the flaw, or the
application is \emph{not affected}, together with the reason why the vulnerable
code cannot be reached or triggered.
VEX-Bench is the most recent work that applies agents for software supply-chain
analysis~\cite{vex-bench}.
However, current agents hallucinate, take shortcuts, and recall facts from
training data instead of analyzing the code at hand.
The original question therefore becomes the one this paper addresses:
\emph{Can we audit the verdict of an agent in software supply-chain analysis?}

A verdict may be incorrect for multiple reasons.
An application marked as \emph{affected} may in fact be safe: agents are trained
to maximize the number of bugs they uncover.
Therefore, an agent may report an unrelated issue discovered along the way, or
declare the application affected on generic grounds, without answering whether
that specific library constitutes a threat.
Conversely, an application marked as \emph{not affected} may in fact be
vulnerable, and for the opposite reason.
Supply-chain analysis frequently ends in a non-problem, because the vulnerable
code is unreachable or the configuration prevents the triggering conditions from
occurring.
Such a verdict is therefore as valuable as its counterpart, yet an agent that
reasons carefully to that conclusion is indistinguishable from an agent that
stops early, misses the path, or invents a mitigating condition.
Current agents are not designed to handle this level of uncertainty, and a bare
verdict carries no information about how the agent reached it.

This is where we position \sys, a framework that addresses the auditable
question by separating the analysis from its validation.
\sys is built around two agents: the \analyst and the \judge.
The \analyst performs the software supply-chain analysis in a monitored
environment, where every tool call, file access, and web request is recorded.
The resulting assessment contains a verdict, the claims supporting that verdict,
and the complete record of the \analyst's interactions with the system.
Then, one or more \judge agents cross-validate the assessment: each \judge
receives the claims and the recorded evidence, and checks whether every claim is
backed by an observed action.
The resulting report scores the assessment along ten dimensions, verifies the
individual claims, and recommends improvements that the \analyst uses in a
further iteration.
The use of dimensions provides a numerical estimation that captures the
\analyst's performance, such as the presence of a working exploit, reachability,
or evidence quality (\autoref{design:judge-report}).
The combination of verdict and dimensions allows \sys to produce an
evidence-backed report that a human analyst audits and acts upon.

We evaluate \sys on a manually constructed dataset of 54 cases covering three
situations: a library vulnerability that is exploitable through the host
application, a library vulnerability that is not exploitable, and a library
without known vulnerabilities.
We run the dataset with both open-weight models (\gemma, \qwen, and \muse) and
frontier models (the GPT-5.6 and Claude 5.5 families), in single-agent and
multi-\judge configurations.
Our results reveal two crucial findings. 
\emph{First, current coding agents, based on frontier or open-weight models, are
still insufficient to perform software supply-chain analysis reliably.}
Without a judge, an \analyst reaches a balanced accuracy between 0.556 (Claude
Opus 5.5) and 0.667 (Qwen 3.8). 
\emph{Second, when combining a panel of low-end and open-weight models, we can boost an open-weight model and surpass the performance of frontier ones.}
The best configuration reaches 0.819 (Qwen 3.8 with a four-judge panel).
From our observations, agents handle host applications whose library is
vulnerable but not exploitable reasonably well, but struggle to declare host
applications without an affected library not affected, as most of them hedge or
report a vulnerability that does not exist.
We consider this behavior the result of model fine-tuning, which optimizes for
bug finding, while software supply-chain analysis often concludes that no issue
exists.
Our evaluation demonstrates that the \judge dimensions, led by reachability and
classification, correlate positively with the correctness of the \analyst
(Spearman $\rho$ up to $+0.60$), thus providing auditable reports for human
analysts.

Our evaluation further documents how agents misbehave in software supply-chain
analysis~(\autoref{sec:agentbehavior}).
For instance, agents cite external evidence that the monitor never records them
retrieving, drift to a newer fixed revision than the one in the prompt,
overstate the completeness of their exploitability chains, and produce proofs of
concept that mock the outcome instead of triggering the vulnerability.

In summary, this paper makes the following contributions:
\begin{itemize}
    \item We observe, characterize, and quantify the failure of agents when
    addressing software supply-chain analysis.

    \item We propose \sys: a \emph{traceable}, \emph{independent}, and \emph{verifiable} framework for software supply-chain analysis.
    
    \item We demonstrate that the audits of \sys can substantially improve 
    the performance of agents during software supply-chain analysis.

    \item We build a dataset of 54 manually curated cases, balanced across
    affected, not affected, and no-known-vulnerability situations.

    \item We evaluate eight models from five vendors, spanning
    open-weight and frontier families, and report how \judge feedback, model
    choice, and analysis cost interact.
\end{itemize}

We release an anonymous prototype of \sys at
\url{https://anonymous.4open.science/r/scarva/README.md}.

\section{Background and Related Work}

\subsection{Common Vulnerabilities and Exposures}
\label{sec:cve}

The Common Vulnerabilities and Exposures (CVE) Program provides unique
identifiers for publicly disclosed cybersecurity vulnerabilities, allowing
advisories, vulnerability databases, and security tools to refer to the same
issue.
A published CVE record describes the vulnerability, identifies affected products
and versions, and links to supporting references~\cite{cve-record-format}.
However, these descriptions often omit precise triggering inputs, execution
conditions, or the environment needed to reproduce the vulnerability.
Consequently, a CVE record alone often lacks the detail needed to test
exploitability directly.

\subsection{Software Bill of Materials}
\label{sec:sbom}

A Software Bill of Materials (SBOM) provides a machine-readable inventory of the
components that make up a software product. 
SBOM reports record component names, versions, suppliers, and dependency
relationships~\cite{ntia-sbom}.
Generators such as Syft~\cite{syft} and Trivy~\cite{trivy} extract this
inventory from a source tree or a container image, while scanners such as
Grype~\cite{grype} and OSV-Scanner~\cite{osvscanner} compare the inventory with
vulnerability databases to identify dependencies associated with known CVEs.
Platforms such as OWASP Dependency-Track~\cite{dependencytrack} collect the
resulting alerts and monitor them over time.
However, a version match alone does not establish whether attacker-controlled
input can reach the vulnerable operation through the host application or whether
its configuration permits exploitation.
An empirical study of SBOM-based scanners attributes most of their false alarms
to exactly this gap, as the flagged vulnerability frequently resides in code
that the application never invokes~\cite{sbom-reality-check}.

\subsection{LLM Agents}

All LLMs we use in this paper are based on the transformer architecture proposed
by Vaswani \etal~\cite{vaswani2017attention}. 
Unlike early LLMs~\cite{radford2018gpt1, brown2020language}, the systems we
focus on have built-in reasoning capabilities.
These models are trained to output reasoning steps that help them solve more
complex tasks like coding and mathematical
problems~\cite{guo2025deepseek,zelikman2022star}.
Furthermore, to enable the models to inspect the code and run commands, we use
agents~\cite{schick2023toolformer,yao2022react}.
These are LLMs that can output special commands (tool calls) that are read by an
interpreter. 
This interpreter can then run terminal commands, fetch files for
the LLM, and run other auxiliary tools.
With the output the model receives from the interpreter, it can decide to make
more tool calls or output a final answer.

Specifically, we use agent harnesses such
as Claude Code and Codex~\cite{anthropic_claude_code,openai_codex}.
They are software frameworks designed to make LLMs perform better at coding
tasks by enabling them to work independently of a human, execute and compile
code, use a web browser, and perform other relevant activities.
Coupled with a harness, LLMs greatly improve their performance on relevant
software engineering tasks~\cite{yang2024swe}.
In particular, they have been shown to be valuable tools for finding security
vulnerabilities~\cite{wang2026cybergym,huggingface2026incident}.
In this work, we focus on the use of these tools for software supply-chain
vulnerabilities. 

\subsection{Related Work}
\label{sec:relwork}

\noindent \textbf{Dependency-Level Analysis.}
A first line of research focuses on pure code analysis, reasoning over code
graphs to decide reachability.
Vuln4Real reduces false alerts by counting only the dependencies that are
actually shipped, separating development-only dependencies, grouping libraries
released as a unit, and flagging halted ones~\cite{vuln4real}.
VulTracer works at function granularity, stitching per-package semantic graphs
to decide whether the vulnerable function is reachable, removing 94\% of the
false positives of \texttt{npm audit}~\cite{vultracer}.
Conceptually, both answer whether a flaw is \emph{present} or \emph{reachable},
which is a necessary precondition for an exploit rather than a proof of one; the
authors of VulTracer leave exploitability out of scope because proving it
requires symbolic execution or taint tracking and is intractable at ecosystem
scale.
\sys takes on that remaining step, including conditions that no call graph
expresses: in \autoref{lst:case-21}, the vulnerable Brotli code is linked into
\texttt{MongoDB} and reachable, yet the default configuration never selects it.
Both approaches also rely on the manifests and resolvable call graphs of Maven
and npm, whereas \sys targets C and C++.

\noindent \textbf{Agentic Security Benchmarks.}
Cybench~\cite{cybench}, ExploitGym~\cite{exploitgym}, and
CyberGym~\cite{wang2026cybergym, cybergym-e2e} measure how well agents discover,
reproduce, and exploit vulnerabilities, and CVE-Genie reproduces CVEs through a
multi-agent framework~\cite{cvegenie}.
Conceptually, all of them score an agent inside a single code base by the
exploits it produces, so a failure to exploit is indistinguishable from a
correct negative.
In contrast, \sys asks a cross-repository question in which the vulnerability is
already known and a \emph{not affected} verdict is a valid answer.

\noindent \textbf{Supply-Chain Exploitability Assessment.}
VEX-Bench is the closest work, evaluating LLM agents on whether an upstream
vulnerability is exploitable downstream, using 75 expert-labeled cases in
Python, Java, and Go~\cite{vex-bench}.
Each agent returns a status, a VEX justification label, and a free-form
reasoning trace, and the evaluation scores the first two against expert labels.
The conceptual difference lies in the validation.
The trace of VEX-Bench is self-reported, as the agent narrates its own steps
without any external check, and may therefore be hallucinated itself.
\sys instead treats trustworthiness as a first-class concern: the actions of the
\analyst are recorded independently of what the \analyst reports, and a separate
\judge checks every claim against that record.
A verdict thus becomes an auditable artifact rather than a label to score.
A second difference is the dataset: every VEX-Bench case starts from a real CVE.
Conversely, our dataset contains a third class of cases, called the control group,
which involves libraries with no known vulnerability and allows us to expose
fabricated findings (\autoref{sec:dataset}).
To the best of our knowledge, \sys is the first to use a control group.

\section{Problem Definition}
\label{sec:probdef}

\begin{figure}[!t]
    \centering
    \resizebox{0.8\columnwidth}{!}{

\begingroup%
\definecolor{pdNavy}{HTML}{001055}%
\definecolor{pdBlue}{HTML}{087EED}%
\definecolor{pdBluePale}{HTML}{E6F3FE}%
\definecolor{pdRed}{HTML}{FA1728}%
\definecolor{pdRedPale}{HTML}{FFE9EB}%
\definecolor{pdGray}{HTML}{8A93A6}%
\definecolor{pdGrayPale}{HTML}{F3F5F9}%
\definecolor{pdSkin}{HTML}{FFD7B5}%
\definecolor{pdHair}{HTML}{3B2A20}%
\definecolor{pdWood}{HTML}{9A6B3F}%

\providecommand{\pdbug}[3]{%
  \begin{scope}[shift={(#1,#2)},scale=#3]%
    \foreach \a in {-1,0,1}{%
      \draw[pdRed,line width=0.9pt] (-0.18,\a*0.17) -- (-0.4,\a*0.17+0.1);%
      \draw[pdRed,line width=0.9pt] ( 0.18,\a*0.17) -- ( 0.4,\a*0.17+0.1);%
    }%
    \draw[pdRed,line width=0.9pt] (-0.07,0.42) -- (-0.17,0.58);%
    \draw[pdRed,line width=0.9pt] ( 0.07,0.42) -- ( 0.17,0.58);%
    \fill[pdRed] (0,-0.02) ellipse (0.21 and 0.33);%
    \fill[pdRed] (0,0.36) circle (0.12);%
    \draw[white,line width=0.6pt] (0,-0.3) -- (0,0.2);%
  \end{scope}%
}%

\begin{tikzpicture}[x=1cm,y=1cm,>=Latex,
  every node/.style={inner sep=0pt,outer sep=0pt,font=\sffamily\fontsize{8.5}{10}\selectfont,text=pdNavy},
  dep/.style={draw=pdGray,line width=0.8pt,rounded corners=3pt,fill=white,minimum width=1.95cm,minimum height=1.0cm,align=center},
  vuln/.style={draw=pdRed,line width=1.4pt,rounded corners=3pt,fill=pdRedPale,minimum width=1.95cm,minimum height=1.0cm,align=center},
  edge/.style={draw=pdGray,line width=0.9pt},
  hot/.style={draw=pdRed,line width=1.3pt,dash pattern=on 3pt off 2pt},
  qmark/.style={circle,draw=pdRed,fill=white,minimum size=0.55cm,font=\sffamily\bfseries\fontsize{10}{11}\selectfont,text=pdRed}]

\useasboundingbox (-0.05,1.8) rectangle (10.15,10.3);

\draw[draw=pdGray,dashed,fill=pdGrayPale,rounded corners=6pt,line width=0.7pt]
  (3.3,3.05) rectangle (10.1,8.3);
\node[anchor=east,font=\sffamily\itshape\fontsize{7.5}{9}\selectfont,text=pdGray]
  at (9.95,3.3) {third-party libraries};

\node[draw=pdBlue,line width=1.4pt,rounded corners=4pt,fill=pdBluePale,minimum width=3.0cm,minimum height=1.2cm,align=center]
  (root) at (6.7,9.6) {{\bfseries\fontsize{11}{13}\selectfont nginx}\\[-1pt]{\fontsize{7.5}{9}\selectfont host application}};

\node[vuln] (ssl)  at (4.45,7.2) {{\bfseries OpenSSL}\\[-1pt]{\fontsize{7.5}{9}\selectfont v1.0.1f}};
\node[dep]  (zlib) at (6.7,7.2)  {{\bfseries zlib}\\[-1pt]{\fontsize{7.5}{9}\selectfont v1.2.8}};
\node[dep]  (pcre) at (8.95,7.2) {{\bfseries PCRE}\\[-1pt]{\fontsize{7.5}{9}\selectfont v8.35}};

\node[vuln] (hb)  at (4.45,4.4) {heartbeat};
\node[dep]  (inf) at (6.7,4.4)  {inflate};
\node[dep]  (rx)  at (8.95,4.4) {regex};
\pdbug{5.45}{4.95}{0.6}

\draw[edge] (root.south) -- ++(0,-0.4) -| (zlib.north);
\draw[edge] (root.south) -- ++(0,-0.4) -| (pcre.north);
\draw[edge] (zlib.south) -- (inf.north);
\draw[edge] (pcre.south) -- (rx.north);

\draw[hot] (root.south) -- ++(0,-0.4) -| (ssl.north);
\draw[hot] (ssl.south) -- (hb.north);
\node[qmark] at (5.6,8.6) {?};
\node[qmark] at (4.45,5.8) {?};

\node[anchor=north,font=\sffamily\bfseries\fontsize{7.5}{9}\selectfont,text=pdRed]
  at (4.45,3.85) {CVE-2014-0160};

\draw[->,line width=1.1pt,draw=pdRed] (10.1,9.6) -- (8.3,9.6);
\node[anchor=south,align=center,font=\sffamily\fontsize{7.5}{9}\selectfont,text=pdRed]
  at (9.3,9.7) {attacker input};

\begin{scope}
\fill[pdNavy] (1.5,6.4) circle (0.5);
\begin{scope}
  \clip (0.3,4.6) rectangle (2.7,5.85);
  \fill[pdNavy] (1.5,4.6) ellipse (1.1 and 1.25);
\end{scope}
\draw[pdWood,line width=2.2pt,line cap=round] (2.45,5.7) -- (2.05,5.25);
\draw[pdNavy,line width=1.4pt,fill=pdBlue!15] (2.7,6.05) circle (0.4);
\draw[white,line width=0.8pt,line cap=round] (2.5,6.15) arc (200:250:0.24);

\draw[pdNavy,line width=0.9pt,fill=white] (1.45,9.0) ellipse (1.4 and 0.62);
\draw[pdNavy,line width=0.9pt,fill=white] (1.55,8.0) circle (0.15);
\draw[pdNavy,line width=0.9pt,fill=white] (1.52,7.6) circle (0.1);
\draw[pdNavy,line width=0.9pt,fill=white] (1.5,7.32) circle (0.06);
\node[font=\sffamily\fontsize{8.5}{10}\selectfont] at (1.45,9.0) {problem or not\textcolor{pdRed}{\bfseries\fontsize{14}{14}\selectfont\,?}};

\end{scope}

\pdbug{3.7}{2.2}{0.5}
\node[anchor=west,font=\sffamily\fontsize{7.5}{9}\selectfont] at (4.1,2.2) {vulnerable (CVE)};
\draw[hot] (6.9,2.2) -- (7.45,2.2);
\node[anchor=west,font=\sffamily\fontsize{7.5}{9}\selectfont] at (7.55,2.2) {reachable?};
\end{tikzpicture}
\endgroup%
}
    \caption{Problem setting. A host application (\texttt{nginx}) depends on a tree of third-party libraries. One of them (\texttt{OpenSSL}) contains a known vulnerability, and the analysis must decide whether the application is actually affected.}
    \label{fig:probdef}
\end{figure}
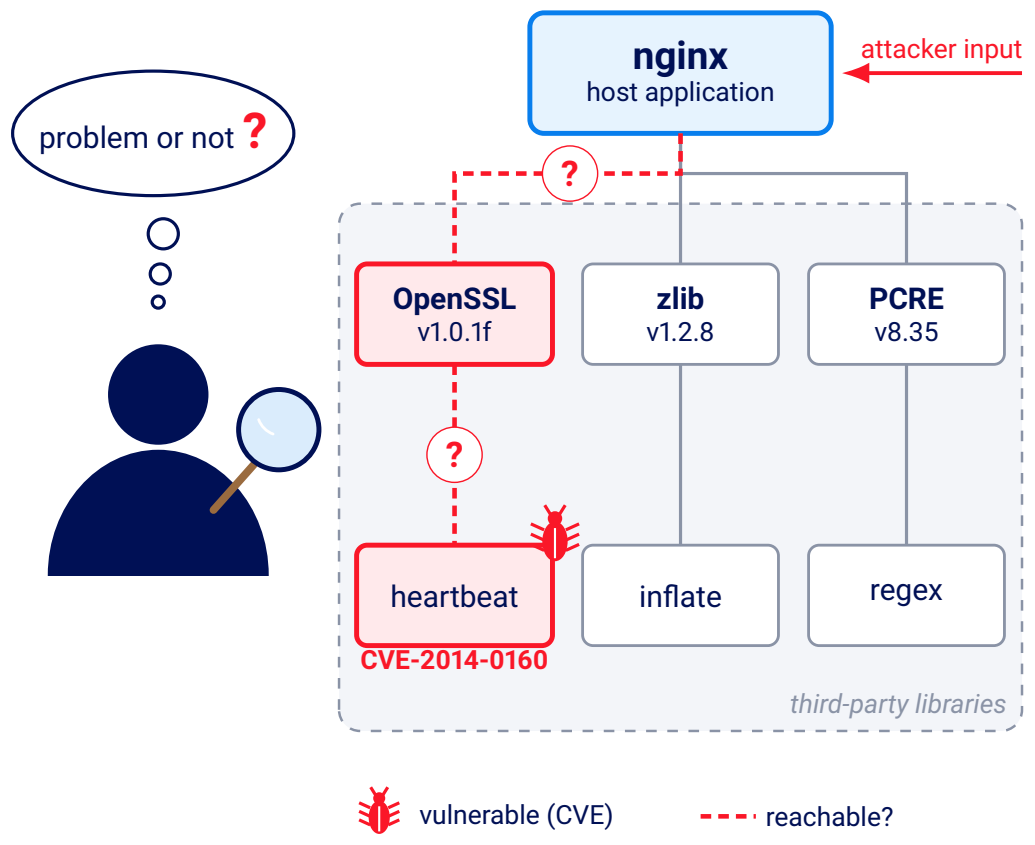

\sys operates in the context of software supply-chain analysis, in which a
vulnerability (CVE, \autoref{sec:cve}) discovered in one of the
\emph{third-party libraries} used by an application may undermine the security
of the application itself.
In this context, a \emph{host application} is a program that exposes accessible
functionality to its users and thereby defines a well-delimited attack surface.
Examples of host applications are web servers (\eg \texttt{nginx}), databases
(\eg \texttt{MongoDB}), and web browsers (\eg \texttt{Chromium}).
In contrast, \emph{third-party libraries} are self-contained pieces of software
that provide specific functionality (\eg \texttt{OpenSSL} exposes cryptographic
primitives) but require a host application to be used.
Note that third-party libraries may themselves depend on other third-party
libraries, defining a software supply chain.
For simplicity, we assume that the vulnerable library interacts directly with
the host application, so the chain has length one.
Even so, our results show that this simplification already presents non-trivial
challenges (\autoref{sec:evaluation}).

Software supply-chain analysis fundamentally differs from classic bug finding,
such as fuzzing or static analysis, which searches for unknown vulnerabilities.
In bug finding, success is measured by the number of bugs uncovered.
Likewise, recent benchmarks for LLM agents ask them to reproduce or exploit
known CVEs~\cite{exploitgym, cybench, cvegenie}.
In contrast, \sys does not aim to maximize the number of findings.
The vulnerability is already known, and the goal is to assess whether it is
exploitable through a specific host application.
Consequently, concluding that the host application is \emph{not affected} is as
valuable as concluding that it is \emph{affected}, and success is measured by the
correctness of the assessment, not by the number of bugs found.

\autoref{fig:probdef} illustrates this setting: \texttt{nginx} is a host
application that includes a list of third-party libraries.
One of them, \texttt{OpenSSL}, contains the Heartbleed vulnerability
(CVE-2014-0160).
Through \texttt{nginx}, an attacker can reach the vulnerable functionality in
\texttt{OpenSSL} and trigger Heartbleed.
Current software supply-chain analysis relies on SBOM tools
(\autoref{sec:sbom}), which analyze the dependency tree.
These tools can recognize whether a dependency is associated with a known CVE,
but they cannot reason about the broader context of the host application and
its attack surface.
Therefore, SBOM tools alone cannot decide whether \texttt{OpenSSL} constitutes a
security threat for \texttt{nginx}.

An adversary can exploit a third-party library if two conditions are met: (i)
there exists an attacker-controllable path connecting the attack surface of the
host application to the vulnerable piece of code, and (ii) a runtime state that
triggers the vulnerability exists.
If both conditions are met, we define the host application as \emph{affected};
otherwise, it is \emph{not affected}.

\begin{listing}[!t]
\begin{lstlisting}[style=ccode]
// File: t1_lib.c
// Function: tls1_process_heartbeat(SSL *s)

/* Read type and payload length first */
if (1 + 2 + 16 > s->s3->rrec.length)
    return 0; /* silently discard */
hbtype = *p++;
n2s(p, payload);
if (1 + 2 + payload + 16 > s->s3->rrec.length)
    /* silently discard per RFC 6520 sec. 4 */
    return 0;
\end{lstlisting}
\caption{Fix of CVE-2014-0160 in OpenSSL v1.0.1g.}
\label{lst:case-01}
\end{listing}

\vspace{0.5em}
\noindent \textbf{Affected Applications:}
\texttt{nginx v1.4.7} uses \texttt{OpenSSL v1.0.1f} as a default dependency,
which contains Heartbleed (\texttt{CVE-2014-0160}).
An adversary can trigger the vulnerability by interacting with an HTTPS endpoint
that accepts attacker-controlled input~\cite{cve20140160,heartbleed}.
\autoref{lst:case-01} shows the fix introduced in the next version,
\texttt{OpenSSL v1.0.1g}: a check on the payload length, which was missing in
the previous version.
Accordingly, the host application is protected by updating the third-party
library to the patched version.

\begin{listing}[!t]
\begin{lstlisting}[style=ccode]
// File: message_compressor_options_server.idl
global:
 cpp_namespace: "mongo"

configs:
  "net.compression.compressors":
   description: "Comma-separated list of compressors to use for network messages"
   source: [cli, ini, yaml]
   arg_vartype: String
   short_name: networkMessageCompressors
   default: "snappy,zstd,zlib" // brotli not used
\end{lstlisting}
\caption{MongoDB compressor options.}
\label{lst:case-21}
\end{listing}

\vspace{0.5em}
\noindent \textbf{Non-Affected Applications:}
\texttt{MongoDB v8.2.6} is an example of a host application that includes a
vulnerable dependency but is \emph{not} affected by it.
The application uses a \texttt{Brotli} version prior to \texttt{v1.0.8}, a
compression library affected by \texttt{CVE-2020-36846}, and can potentially
reach the flaw.
However, the default configuration does not enable Brotli for network messages,
as shown in \autoref{lst:case-21}.
Therefore, there is no attacker-controlled path to the vulnerable code
location.

\vspace{0.5em}
\noindent \textbf{Non-Exploitability.}
Proving non-exploitability is considered an unsolvable problem in security and
is not the goal of \sys.
In software supply-chain analysis, the practical approach is to analyze the
attack surface and to draw evidence-backed conclusions on whether a specific
condition constitutes a problem.
Accordingly, \sys provides a list of evidence that explains why a CVE is not
exploitable, such as a fixed library version, an unreachable operation, or an
unmet precondition.
These conclusions are paramount for a human analyst to decide the best course of
action.
Therefore, in case of a \emph{not affected} verdict, \sys must provide trustworthy support
for the follow-up actions.

\section{\sys}
\label{sec:design}

\sys performs software supply-chain analysis and assesses whether a vulnerable
third-party library affects or does not affect the host application
(\autoref{sec:probdef}).
To achieve this goal, the system must go beyond a simple verdict (\eg
\emph{affected} or \emph{not affected}), because a bare verdict may be caused by
shortcuts or hallucinations.
The system must also provide verifiable evidence supporting its claims.
\sys addresses this issue through a verifiable report that summarizes the
reasons behind a verdict and ties the supporting evidence to concrete program
statements or documentation.
This mirrors the work of a human analyst, transposed to LLM agents.

The particular needs of software supply-chain analysis motivate the three main
principles of \sys: \emph{traceable} (the security analysis is recorded),
\emph{independent} (the verdict is assessed separately), and \emph{verifiable}
(every claim is rooted in evidence).
\sys adheres to these principles through three design choices.
First, an \analyst agent performs the software analysis, and its actions are
\emph{traced}.
Second, one or more \judge agents \emph{independently} check the \analyst's
results.
Third, \sys produces a \emph{verifiable} report that contains the verified
claims about the entire analysis process.
Separating these roles matters because an agent that grades its own work tends
to confirm it, so the \judge works only from the \analyst's claims and the
recorded evidence.

\begin{figure*}[!t]
    \centering
    \makebox[\textwidth][c]{%
        \hspace*{-20pt}
        \resizebox{\textwidth}{!}{\input{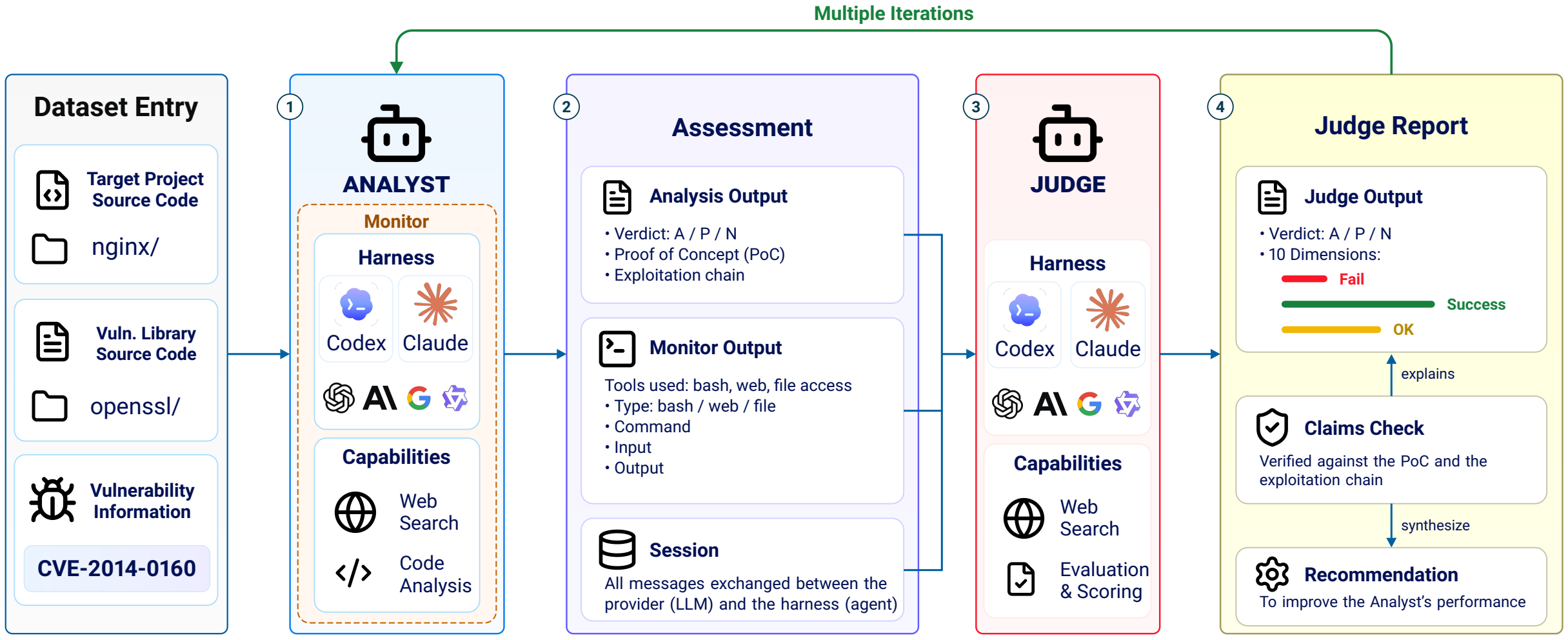}}
        \hspace*{20pt}
    }
    \caption{\sys pipeline with the data structs exchanged between \analyst, \judge, and the final report.}
    \label{fig:design}
\end{figure*}

\autoref{fig:design} shows the \sys pipeline, which is composed of four components
numbered \designcircnum{1}--\designcircnum{4}, and the same numbers mark the
corresponding subsections (\autoref{design:analyst}--\ref{design:judge-report}).
A \sys dataset entry consists of the source code of the host application, the
library, and the vulnerability information.
The \analyst~\designcircnum{1} thus examines the dataset entry and verifies (or
disproves) whether the library affects the security of the application.
The \analyst runs in a monitored environment where every tool call, including
web searches and file access, is logged.
The \analyst's output is an assessment~\designcircnum{2} containing a verdict
(\emph{affected}, \emph{not affected}, or \emph{potentially affected}), a list
of claims supporting the verdict, and all the monitor data collected during the analysis.
For the verdict, we include a \emph{potentially affected} case to capture
the agent's uncertainty; we discuss the implication of this choice in the
\analyst section (\autoref{design:analyst}) and the evaluation
(\autoref{sec:evaluation}).
The claims, combined with the observed interactions, allow an independent
evaluation of the \analyst's work.
The assessment is then handed off to the \judge~\designcircnum{3}, which
cross-validates the \analyst's conclusions by checking whether each claim is
supported by the monitor data.
To reduce the bias of any single model, \sys can run one or more independent
\judge agents, each using a different LLM.
Each \judge produces a report~\designcircnum{4} that scores how trustworthy the
\analyst's assessment is, verifies its claims through claim checks, and
provides recommendations to improve the \analyst's performance.

The \judge reports can optionally be returned to the \analyst as feedback for
further iterations, in which the \analyst revisits the same case and produces a
revised assessment.
The reports remain separate outputs, so each piece of feedback can be traced to
the \judge model that produced it.
In our experiments, this feedback helps the \analyst reduce hallucinations and
improve its performance (\autoref{sec:evaluation}).

\sys's final output consists of the \analyst's assessment and the \judge's
report, which together support human operators in taking action, for example by
updating the vulnerable library or developing a patch.
In the following, we describe each component in detail.

\subsection{\analyst \protect\designcircnum{1}}
\label{design:analyst}

The \analyst agent performs the vulnerability assessment and is based on the
ReAct paradigm~\cite{yao2022react}.
The main goal of the \analyst is to determine whether the host application is
exploitable through a vulnerability in a library it uses.
The \analyst examines the source code of both the library and the host
application to identify a potential exploit chain and the conditions required to
exploit it.
The assessment is performed without a live system, and is based on the source
code.
However, the agent can build and interact with the project through regular shell
commands.
A case is classified as \emph{affected} when the vulnerability is exploitable
in the host application, and as \emph{not affected} when it is not exploitable
through the host application or the library is not vulnerable.
A case is classified as \emph{potentially affected} when the evidence is
insufficient to support either conclusion, which avoids a forced binary
decision.
In practical terms, a human analyst may decide to behave conservatively, thus
considering \emph{potentially affected} as \emph{affected}, or in a more relaxed
way, thus considering \emph{potentially affected} as \emph{not affected}.
In our evaluation, we measure the impact of different policies that bucket
\emph{potentially affected} as \emph{affected} or \emph{not affected},
respectively (\autoref{sec:evaluation}).

\noindent \textbf{Capabilities:}
The \analyst can use web search and web scraping, and can access the source code
of the library and the host application, both provided as Git repositories.
A Python script, provided as a skill, supports creation and validation of the
report.
Moreover, the agent has free access to common bash commands and can install
additional tools if needed.

\noindent \textbf{Output:}
The \analyst's output consists of (1) a final report containing the verdict and
the supporting claims, (2) a proof of concept (PoC) for affected claims, and (3)
a list of concrete steps to trigger the exploit.
Note that (2) and (3) are not mandatory.
In our evaluation, we observe that the \analyst rarely provides a working PoC,
yet it provides a correct exploitation description backed by valid evidence.
Leaving the PoC optional also tests whether the LLM fabricates an
\emph{affected} verdict when no exploit is possible.

\subsection{Assessment \protect\designcircnum{2}}
\label{design:assessment}

The assessment comprises the \analyst's output, the monitor data, and the
session, \ie all the messages exchanged between the harness and the model
provider.
The \analyst's output is considered untrusted, as it may be hallucinated by the LLM.
Conversely, the monitor data and session are considered trusted as they are
collected independently.
This structure allows the \judge agent to verify if a claim is backed by evidence.
For instance, the \analyst may claim that a source file was inspected, but the
monitor data do not show any file access.
Claims without corresponding evidence may reflect unsupported reasoning or
information recalled from training data.
Consistency between claims and collected evidence is therefore a central part of
the assessment.

\subsection{\judge \protect\designcircnum{3}}
\label{design:judge}

The \judge agent evaluates the quality of the \analyst's assessment and, like
the \analyst, is based on the ReAct paradigm~\cite{yao2022react}.
By design, the \judge has access only to the assessment, and not to the source
code of the project or the library.
This design choice prevents the \judge from replicating the \analyst's security
assessment, and lets it focus on the evidence contained in the monitor data
and the session.
The verdict of the \judge is therefore a re-assessment based solely on the
evidence in the assessment, and not on its own analysis of the source code.
The \judge has two purposes.
First, it scores the assessment across ten dimensions, which reflect our
empirical observations on the main reasons an \analyst may incorrectly conduct a
software supply-chain analysis (see \autoref{design:judge-report}).
Second, alongside the dimensions, the \judge produces a list of claim checks, \ie
independent verifications of the analysis output.
The claim checks explain the dimensions and allow the \judge to produce a list
of recommendations that help the \analyst improve its assessment.

When multiple \judge agents evaluate an assessment, each runs in a separate
container and evaluates independently, without access to the other \judge
agents' evaluations.

\noindent \textbf{Capabilities:}
The \judge can use web search and web scraping, and
receives the \analyst's assessment.
The \judge cannot access the source code of the host application or the library
directly, except for code included in the assessment.

\noindent \textbf{Output:}
The \judge report contains (1) a re-assessed verdict in the same three classes
as the \analyst, (2) a score per dimension from zero (fail) to two (success),
(3) claim checks, and (4) recommendations for further iterations.

\subsection{Judge Report \protect\designcircnum{4}}
\label{design:judge-report}

The judge report includes four parts: a re-assessed verdict, a scoreboard of ten
dimensions, a list of claim checks, and a list of recommendations for the
\analyst agent.

The verdict is the \judge's own re-assessment of the case, chosen among
\emph{affected}, \emph{not affected}, and \emph{potentially affected}, based on
the evidence in the submitted assessment.
The verdict can therefore confirm or contradict the \analyst's conclusion.

A \emph{dimension} is an independent criterion for assessing one specific aspect
of the \analyst's work, phrased as a question (\eg ``Did the \analyst examine the
correct package?'').
Together, the ten dimensions capture how complete and well-supported each part of
the \analyst's process is, and thus measure how much the \analyst can be
trusted.
We design them after observing how the \analyst agent performs software
supply-chain analyses.
Each dimension receives a score from \emph{zero} to \emph{two} that quantifies
how much the \judge trusts the \analyst on that aspect.
We choose a three-value scale instead of a binary one to make the scoring
more stable, following previous work~\cite{lee2025consistency}.
The dimension scores are summed to a maximum of twenty.
A high total indicates a high level of trust in the \analyst, and therefore in
its verdict.

Along with the dimensions, the \judge reports a list of claim checks, which
explain in detail why each dimension received its score.
The claim checks also give a human analyst the evidence needed to support their
decision process.
Finally, the claim checks are distilled into a list of recommendations that help
the \analyst avoid mistakes in further iterations.

\noindent \textbf{Dimensions:}
The dimensions fall into four groups.
Scope, dependency, and vulnerability represent the \analyst's capability to
identify the relevant code and vulnerable behavior.
Integration and reachability capture whether the \analyst's assessment is
compatible with the host application.
Evidence, tool use, and PoC address the support for the \analyst's claims.
Classification and artifacts address the conclusion and the required
deliverables.

The ten dimensions are listed below:

\begin{itemize}
    \item \textbf{Scope:} Whether the \analyst examines the correct target package, dependency path, and relevant source files.
    \item \textbf{Dependency:} Whether the \analyst uses the case-specific
    dependency versions, commits, tags, and ranges.
    \item \textbf{Vulnerability:} Whether the dependency version or code is on the vulnerable or fixed side of the CVE boundary, and why.
    \item \textbf{Integration:} How the host application connects attacker-controlled input to the vulnerable library API.
    \item \textbf{Reachability:} The runtime, build, feature, platform, and configuration conditions required for the integrated path to execute.
    \item \textbf{Evidence:} The quality and relevance of supporting material, such as local source code, build metadata, authoritative advisories, patches, or PoC results.
    \item \textbf{Tool Use:} Whether recorded tool calls support the claimed inspection or validation.
    \item \textbf{Classification:} Whether the \analyst's final verdict (affected, not affected, or potentially
    affected) is supported by the evidence and consistent across its output.
    \item \textbf{PoC:} Whether the PoC's presence, design, execution result, and reported status are accurate.
    \item \textbf{Artifacts:} Whether the \analyst produced the required output files,
    including the conditional PoC files that support the verdict.
\end{itemize}

\section{Implementation}

\label{sec:implementation}

\sys consists of several modules: a pipeline orchestration interface
(\autoref{impl:pipeline}), containers equipped for running the \analyst and the
\judge (\autoref{impl:runtime}), auditing for monitoring the agents' progress
(\autoref{impl:auditing}), and LLM access points for open-weight and proprietary
models (\autoref{impl:llm}).

\subsection{Pipeline Orchestration}
\label{impl:pipeline}

The pipeline of \sys is orchestrated by a command-line tool (CLI) written in
Python that selects the harness, the Docker image~\cite{docker}, and the case
repositories supplied through command-line flags and arguments.

Following \autoref{sec:design}, the CLI starts the \analyst, which produces
three output objects.
A JSON report contains the \analyst's structured verdict, Markdown notes
describe the model's reasoning, and a Proof of Concept (PoC) demonstrates the
exploit for \emph{affected} or \emph{potentially affected} cases.
The CLI passes these items, together with the tool-call history, to the \judge,
which produces a \judge report.
Reports from different \judge agents are kept separate.
Most importantly, the \judge agents do not know which model produced the
analysis, thus avoiding implicit bias.
The second iteration starts a new session and workspace and repeats the analysis
using the corresponding reports as feedback.
The second \analyst run is then evaluated by the \judge.

\subsection{Agent Runtime}
\label{impl:runtime}

The agent runtimes for the \analyst and the \judge are containerized via Docker.
Containerization ensures an identical setup across runs and confines the agents.
Prebuilt images provide either Codex, in version v0.147.0 for OpenAI models and
v0.116.0 for open-weight models, or Claude Code version
v2.1.293~\cite{openai_codex, anthropic_claude_code}.
Newer versions of Codex contain a bug that prevents non-GPT models from handling
tool calls properly~\cite{openai_codex_bug}.
Each agent runs in a separate container, and no state is carried between runs.
The containers execute agents as unprivileged users, but agents can raise their
privilege level inside the container on demand (\ie the account belongs to
\texttt{sudoers}).

Both the \analyst and the \judge have access to web search via Firecrawl
and browser interaction via a Playwright Model Context Protocol (MCP)
server~\cite{firecrawl_mcp, playwright_mcp, modelcontextprotocol}.
They also have access to skills that explain how the tools should be used,
ensure the agent runs until it completes the task, and validate the report
passed to the next stage of the pipeline.

The \analyst gets a Markdown file with the prompt detailing its task
(Appendix~\ref{app:prompts}).
The \analyst also gets the CVE name and the target and library repositories,
mounted as read-only to prevent modifications to the supplied source code across
assessments.
Moreover, the \analyst can write its analysis and notes only in designated
directories.
To ensure the report JSON is well formed, the \analyst validates the report with
a Python script provided as a skill.
On the second iteration, the \analyst also gets the reports from the \judge,
together with a second prompt file explaining where to find the output of the
\judge (Appendix~\ref{app:prompts}).

The \judge likewise receives a Markdown file with the prompt that explains its
task (Appendix~\ref{app:prompts}), together with read-only mounts of the
\analyst's output workspace, metadata, and monitored tool calls.
As for the \analyst, a Python script validates the JSON output format.
Neither the source repositories nor logs are mounted into the \judge container,
which is instructed to treat its inputs as untrusted and never execute them.

\subsection{Auditing}
\label{impl:auditing}

All agents connect to a Docker Compose environment that provides monitoring.
Both harnesses emit OpenTelemetry messages, which are stored by
Loki~\cite{opentelemetry, grafana_loki}, while a Grafana dashboard displays the
recorded activity in real time~\cite{grafana}.

Tool-call records provide evidence of tool-mediated file access and
web requests.
Loki provides the Codex tool calls, while Claude Code writes them to a file that
is processed by the CLI.
These records are later used by the \judge to evaluate the \analyst.

After the analysis, the CLI exports tool names, arguments, outputs, and backend
identifiers into one JSON file.
If the output file is missing or invalid, the \judge is not called.
The CLI also warns about invalid reports produced by the agent, which indicate
either a technical issue or a request aborted by cyber-guardrails.
In the case of a technical issue, the assessment is discarded and repeated.

\subsection{LLM Access Points}
\label{impl:llm}

Open-weight LLMs without quantization are used alongside provider-hosted OpenAI
and Anthropic models.
To host the open-weight models, vLLM v0.28.0 instances are deployed on a server with
four NVIDIA H100 GPUs, 1TB of RAM, and 96 CPU cores.
Each instance serves one model through an OpenAI- or Anthropic-compatible API,
with multiple instances used concurrently.
Each model is hosted with the default settings derived from the standard
configuration of the model provider, combined with the default hosting settings
of vLLM.
In addition, all models use their highest available thinking effort level, and
their context windows are left unchanged.

The OpenAI and Anthropic models are queried via API.
The Anthropic account used is enrolled in the Cyber Verification
Program.\footnote{\url{https://www.anthropic.com/news/cyber-verification-program}}
For OpenAI, we did not obtain the equivalent verification.
This limitation lets us observe agent misbehavior caused by such restrictions
(\autoref{sec:agentbehavior}).

\section{Evaluation}
\label{sec:evaluation}

The evaluation of \sys is designed to address the following questions.

\begin{itemize}
    \item \textbf{RQ1:} How does \sys perform in software supply-chain
    analysis? (\autoref{sec:perform})
    \item \textbf{RQ2:} How reliable is the report produced by the \judge?
    (\autoref{sec:trust})
    \item \textbf{RQ3:} How does \sys perform compared with VEX-Bench?
    (\autoref{sec:vexcomp})
    \item \textbf{RQ4:} How do the agents behave when performing software
    supply-chain analyses? (\autoref{sec:agentbehavior})
\end{itemize}

\begin{table}[t]
    \centering
    \captionsetup{width=\columnwidth}
    \caption{The evaluated models. Closed-weight vendors do not disclose the model size. Muse Glimmer acts only as \judge.}
    \label{tab:model-cutoffs}
    \footnotesize
    \setlength{\tabcolsep}{3.5pt}
    \renewcommand{\arraystretch}{1.1}
    \begin{tabular}{@{}l l r l@{}}
        \toprule
        \textbf{Model} & \textbf{Vendor} & \textbf{Params} & \textbf{Cutoff} \\
        \midrule
        \quad \gemma~\cite{gemmateam2026gemma4} & Google & 31B & Jan. 2025 \\
        \quad \qwen~\cite{qwen38} & Alibaba & 27B & Apr. 2025~\cite{benchgecko_knowledge_horizon} \\
        \quad \muse~\cite{museglimmerteam2026museglimmer30b} & Meta & 30B & Jan. 4, 2026 \\
        \midrule
        \quad GPT-5.6 family~\cite{openai2026gpt56} & OpenAI & --- & Feb. 16, 2026 \\
        \quad Claude 5.5 family~\cite{anthropic2026opus55, anthropic2026sonnet55} & Anthropic & --- & Jun. 2026 \\
        \bottomrule
    \end{tabular}
\end{table}

\subsection{Dataset}
\label{sec:dataset}

The dataset contains 54 cases, divided into three groups:
\begin{itemize}
    \item \dg{1} contains known library vulnerabilities that are exploitable
    through the target program (18 cases) and is labeled as \emph{affected}.
    \item \dg{2} contains known library vulnerabilities that are not
    exploitable through the target program (18 cases) and is labeled as
    \emph{not affected}.
    \item \dg{3} contains no known vulnerabilities in its library (18 cases) and
    is labeled as \emph{not affected}.
\end{itemize}

The vulnerabilities used in \dg{1} and \dg{2} are listed in
\autoref{tab:scarva-cves} (Appendix~\ref{app:dataset}).
All the cases are selected and constructed through manual research, and the
ground-truth labels are based on known exploits.
The groups separate two questions: whether the library contains a known
vulnerability and whether the target program can reach and exploit it. 
A dataset entry given to the \analyst contains the source code of the target
program, the library, and the vulnerability information.
As only \dg{1} contains vulnerabilities with a known exploit, only cases in this
group are labeled as \emph{affected}, while \dg{2} cases are labeled
\emph{not affected}.
The third group, \dg{3}, serves as a negative control to evaluate whether agents
correctly recognize the absence of known vulnerabilities rather than reporting
unrelated or fabricated findings. 
By their nature, \dg{3} cases are labeled \emph{not affected}.
The last group is particularly important because agents optimized for
vulnerability discovery may incorrectly classify applications as affected under
uncertainty.
\sys is the first work on agentic software supply-chain analysis to include a
control group without any known vulnerability (\dg{3}), while previous works only
consider \emph{affected} and \emph{not affected} cases that start from a known
CVE~\cite{vex-bench}.

Each case indicates the affected application and library versions, the
repositories' source code, the conditions for the assessment, the assessment
label, and the rationale.
Exploit descriptions and PoC URLs are included when available. This information
is used only for manual verification and is leaked neither to the \analyst nor
to the \judge. 
For example, the case of \texttt{CVE-2014-0160}, \texttt{nginx 1.4.7} with
\texttt{OpenSSL 1.0.1f}, is labeled \emph{affected} because the original
Windows binary uses the bundled OpenSSL release with HTTPS and heartbeat
support. 
The case rationale describes Heartbleed's missing bounds check, which allows
a remote memory over-read~\cite{cve20140160,heartbleed}. 
The case of \texttt{CVE-2020-36846}, \texttt{MongoDB} server \texttt{8.2.6} with
\texttt{Brotli} below \texttt{1.0.8}, is labeled \emph{not affected} under the
standard MongoDB compression configuration because that scenario does not use
the one-shot Brotli decompression path~\cite{cve202036846}.

Some selected vulnerabilities, including Heartbleed, are sufficiently old that
they may appear in the models' training data.
Results on these cases may therefore reflect prior model knowledge as well as
reasoning over the supplied repositories and evidence. 
The benchmark should be interpreted with this possible overlap in mind.

Since \sys can issue \emph{potentially affected} verdicts, we employ three
policies to translate this verdict to our dataset labels: \emph{conservative},
\emph{relaxed}, and \emph{excluded}.
The \emph{conservative} policy treats \emph{potentially affected} cases as
\emph{affected}, the \emph{relaxed} policy treats the \emph{potentially
affected} as \emph{not affected}, and the \emph{excluded} policy simply ignores
the \emph{potentially affected} cases entirely.
The following experiments use the \emph{conservative} policy, since we consider
this scenario the most realistic one, \ie a human analyst wants to take action
in the case of a suspected threat.
We further measure the impact of the other policies in Appendix~\ref{app:policies}.

\subsection{Configuration}
\label{ssec:exp-config}

This section describes the \sys configurations used in our evaluation.
Table~\ref{tab:model-cutoffs} lists the evaluated models, their vendors and
parameter sizes, and their stated training-data cutoffs.
A cutoff is the latest date of data collected for training.
We run every \analyst agent in three settings: without a \judge (\emph{None}),
with the same \analyst model as \judge (\emph{Self}), and with a low-power model
from a different vendor as \judge (\emph{Luna} for \luna or \emph{Haiku} for
Claude \haiku).
For the two open-weight \analyst models, \gemma and \qwen, we further test a
panel of four \judge agents composed of \qwen, \gemma, \luna, and \muse
(\emph{Panel}).
In every configuration with a \judge, the \analyst receives the \judge's report
and runs a second iteration, which lets us measure whether the \judge feedback
improves its performance.
We run every configuration three times and report the mean and the standard
deviation over the runs.
This design lets us measure the impact of a \judge on different \analyst models.
More importantly, the different \judge configurations reveal how the performance
of the \analyst changes with the feedback.
The following sections define the metrics and discuss our observations for each
research question (\autoref{sec:perform} to \autoref{sec:agentbehavior}).

\subsection{\sys Performance (RQ1)}
\label{sec:perform}

\begin{table*}[t]
    \centering
    \captionsetup{width=\textwidth}
    \caption{Balanced accuracy and per-group recall (mean $\pm$ SD over runs) by
    \analyst and \judge configuration. $\Delta$ is the change in balanced
    accuracy relative to the same \analyst without a \judge. Bold marks the best
    value per column. \dg{1}: exploitable library vulnerability; \dg{2}: library
    vulnerability that is not exploitable; \dg{3}: library without known
    vulnerability.}
    \label{tab:results}
    \small
    \setlength{\tabcolsep}{15pt}
    \renewcommand{\arraystretch}{1.1}
    \begin{tabular}{ll r r ccc}
        \toprule
        & & \multicolumn{2}{c}{\textbf{Balanced accuracy}} & \multicolumn{3}{c}{\textbf{Recall by dataset group}} \\
        \cmidrule(lr){3-4}\cmidrule(lr){5-7}
        \textbf{\analyst} & \textbf{\judge} & & $\Delta$ & \dg{1} & \dg{2} & \dg{3} \\
        \midrule
        \gemma & None & 0.583{\scriptsize\,$\pm$\,0.024} & \textcolor{black!50}{---} & 0.352{\scriptsize\,$\pm$\,0.032} & 0.870{\scriptsize\,$\pm$\,0.032} & 0.759{\scriptsize\,$\pm$\,0.085} \\
         \rowcolor{gray!12} & Self & 0.597{\scriptsize\,$\pm$\,0.014} & \textcolor{deltapos}{+0.014} & 0.333{\scriptsize\,$\pm$\,0.000} & 0.870{\scriptsize\,$\pm$\,0.032} & 0.852{\scriptsize\,$\pm$\,0.085} \\
         & Luna & 0.579{\scriptsize\,$\pm$\,0.029} & \textcolor{deltaneg}{$-$0.005} & 0.333{\scriptsize\,$\pm$\,0.000} & 0.889{\scriptsize\,$\pm$\,0.056} & 0.759{\scriptsize\,$\pm$\,0.085} \\
         \rowcolor{gray!12} & Panel$^{\dagger}$ & 0.653{\scriptsize\,$\pm$\,0.000} & \textcolor{deltapos}{+0.069} & 0.389{\scriptsize\,$\pm$\,0.000} & 0.907{\scriptsize\,$\pm$\,0.032} & \textbf{0.926}{\scriptsize\,$\pm$\,0.032} \\
        \midrule
        \qwen & None & 0.667{\scriptsize\,$\pm$\,0.024} & \textcolor{black!50}{---} & 0.574{\scriptsize\,$\pm$\,0.064} & 0.963{\scriptsize\,$\pm$\,0.064} & 0.556{\scriptsize\,$\pm$\,0.056} \\
         \rowcolor{gray!12} & Self & 0.708{\scriptsize\,$\pm$\,0.028} & \textcolor{deltapos}{+0.042} & 0.574{\scriptsize\,$\pm$\,0.064} & 0.963{\scriptsize\,$\pm$\,0.064} & 0.722{\scriptsize\,$\pm$\,0.000} \\
         & Luna & 0.782{\scriptsize\,$\pm$\,0.029} & \textcolor{deltapos}{+0.116} & 0.796{\scriptsize\,$\pm$\,0.064} & \textbf{1.000}{\scriptsize\,$\pm$\,0.000} & 0.537{\scriptsize\,$\pm$\,0.032} \\
         \rowcolor{gray!12} & Panel$^{\dagger}$ & \textbf{0.819}{\scriptsize\,$\pm$\,0.064} & \textcolor{deltapos}{+0.153} & 0.796{\scriptsize\,$\pm$\,0.116} & \textbf{1.000}{\scriptsize\,$\pm$\,0.000} & 0.685{\scriptsize\,$\pm$\,0.032} \\
        \midrule
        \luna & None & 0.597{\scriptsize\,$\pm$\,0.028} & \textcolor{black!50}{---} & 0.741{\scriptsize\,$\pm$\,0.032} & 0.815{\scriptsize\,$\pm$\,0.032} & 0.093{\scriptsize\,$\pm$\,0.085} \\
         \rowcolor{gray!12} & Self & 0.579{\scriptsize\,$\pm$\,0.080} & \textcolor{deltaneg}{$-$0.019} & 0.722{\scriptsize\,$\pm$\,0.096} & 0.704{\scriptsize\,$\pm$\,0.032} & 0.167{\scriptsize\,$\pm$\,0.096} \\
         & Haiku & 0.611{\scriptsize\,$\pm$\,0.000} & \textcolor{deltapos}{+0.014} & 0.759{\scriptsize\,$\pm$\,0.032} & 0.722{\scriptsize\,$\pm$\,0.056} & 0.204{\scriptsize\,$\pm$\,0.032} \\
        \midrule
        \rowcolor{gray!12} \terra & None & 0.592{\scriptsize\,$\pm$\,0.043} & \textcolor{black!50}{---} & 0.623{\scriptsize\,$\pm$\,0.066} & 0.782{\scriptsize\,$\pm$\,0.053} & 0.337{\scriptsize\,$\pm$\,0.073} \\
         & Self & 0.562{\scriptsize\,$\pm$\,0.054} & \textcolor{deltaneg}{$-$0.030} & 0.589{\scriptsize\,$\pm$\,0.084} & 0.786{\scriptsize\,$\pm$\,0.056} & 0.273{\scriptsize\,$\pm$\,0.122} \\
         \rowcolor{gray!12} & Haiku & 0.617{\scriptsize\,$\pm$\,0.027} & \textcolor{deltapos}{+0.025} & 0.685{\scriptsize\,$\pm$\,0.032} & 0.833{\scriptsize\,$\pm$\,0.056} & 0.252{\scriptsize\,$\pm$\,0.094} \\
        \midrule
        \sol & None & 0.576{\scriptsize\,$\pm$\,0.058} & \textcolor{black!50}{---} & 0.685{\scriptsize\,$\pm$\,0.073} & 0.660{\scriptsize\,$\pm$\,0.011} & 0.126{\scriptsize\,$\pm$\,0.016} \\
         \rowcolor{gray!12} & Self & 0.601{\scriptsize\,$\pm$\,0.045} & \textcolor{deltapos}{+0.025} & 0.709{\scriptsize\,$\pm$\,0.096} & 0.614{\scriptsize\,$\pm$\,0.074} & 0.258{\scriptsize\,$\pm$\,0.013} \\
         & Haiku & 0.557{\scriptsize\,$\pm$\,0.022} & \textcolor{deltaneg}{$-$0.019} & 0.723{\scriptsize\,$\pm$\,0.029} & 0.593{\scriptsize\,$\pm$\,0.032} & 0.151{\scriptsize\,$\pm$\,0.014} \\
        \midrule
        \rowcolor{gray!12} Claude \haiku & None & 0.579{\scriptsize\,$\pm$\,0.056} & \textcolor{black!50}{---} & 0.704{\scriptsize\,$\pm$\,0.085} & 0.759{\scriptsize\,$\pm$\,0.064} & 0.148{\scriptsize\,$\pm$\,0.032} \\
         & Self & 0.602{\scriptsize\,$\pm$\,0.032} & \textcolor{deltapos}{+0.023} & 0.796{\scriptsize\,$\pm$\,0.032} & 0.648{\scriptsize\,$\pm$\,0.064} & 0.167{\scriptsize\,$\pm$\,0.000} \\
         \rowcolor{gray!12} & Luna & 0.611{\scriptsize\,$\pm$\,0.024} & \textcolor{deltapos}{+0.032} & \textbf{0.815}{\scriptsize\,$\pm$\,0.032} & 0.667{\scriptsize\,$\pm$\,0.000} & 0.148{\scriptsize\,$\pm$\,0.032} \\
        \midrule
        Claude \sonnet & None & 0.593{\scriptsize\,$\pm$\,0.021} & \textcolor{black!50}{---} & 0.611{\scriptsize\,$\pm$\,0.000} & 0.889{\scriptsize\,$\pm$\,0.056} & 0.259{\scriptsize\,$\pm$\,0.032} \\
         \rowcolor{gray!12} & Self & 0.754{\scriptsize\,$\pm$\,0.050} & \textcolor{deltapos}{+0.161} & 0.667{\scriptsize\,$\pm$\,0.056} & 0.833{\scriptsize\,$\pm$\,0.056} & 0.849{\scriptsize\,$\pm$\,0.086} \\
         & Luna & 0.562{\scriptsize\,$\pm$\,0.018} & \textcolor{deltaneg}{$-$0.030} & 0.648{\scriptsize\,$\pm$\,0.032} & 0.778{\scriptsize\,$\pm$\,0.000} & 0.170{\scriptsize\,$\pm$\,0.056} \\
        \midrule
        \rowcolor{gray!12} Claude \opus & None & 0.556{\scriptsize\,$\pm$\,0.050} & \textcolor{black!50}{---} & 0.537{\scriptsize\,$\pm$\,0.032} & 0.889{\scriptsize\,$\pm$\,0.000} & 0.259{\scriptsize\,$\pm$\,0.140} \\
         & Self & 0.630{\scriptsize\,$\pm$\,0.018} & \textcolor{deltapos}{+0.075} & 0.518{\scriptsize\,$\pm$\,0.063} & 0.889{\scriptsize\,$\pm$\,0.000} & 0.586{\scriptsize\,$\pm$\,0.166} \\
         \rowcolor{gray!12} & Luna & 0.612{\scriptsize\,$\pm$\,0.074} & \textcolor{deltapos}{+0.057} & 0.729{\scriptsize\,$\pm$\,0.105} & 0.852{\scriptsize\,$\pm$\,0.064} & 0.132{\scriptsize\,$\pm$\,0.030} \\
        \bottomrule
    \end{tabular}
    \par\smallskip
    \noindent\begin{minipage}{\textwidth}\footnotesize\raggedright
        \emph{None}: no \judge (baseline). \emph{Self}: the \analyst model also acts as \judge. $^{\dagger}$Panel of \judge{s}: \qwen, \gemma, \luna, and \muse.
    \end{minipage}
\end{table*}

This section studies the performance of \sys under different configurations.
Since the dataset has three equal groups (\dg{1}, \dg{2}, and \dg{3}) but only
two verdict classes (\autoref{sec:dataset}), we measure the performance of the \analyst by
its balanced accuracy.
Moreover, we report the recall within each group, as it shows where an \analyst
fails.
\autoref{tab:results} lists the results for every \analyst without a \judge and
with each \judge configuration.
As stated in \autoref{sec:dataset}, our experiments are executed by adopting a
conservative policy that merges \emph{potentially affected} and \emph{affected},
while in Appendix~\ref{app:policies} we report figures with alternative policies.
Appendix~\ref{app:verdict-breakdown} illustrates the performance at
case granularity.

\paragraph*{Overall Results}
The best-performing \analyst model on our benchmark is \qwen with multiple
\judge models, at 81.9\% balanced accuracy.
Conversely, the worst-performing \analyst model is Claude \opus without a \judge
at 55.6\% balanced accuracy.
The best-performing frontier \analyst model is Claude \sonnet with itself as
\judge at 75.4\% balanced accuracy.
The cheapest frontier models, \luna and Claude \haiku, peak at 61.1\%,
which is slightly better than \gemma at 58.3\% without a \judge or 59.7\% with
itself as \judge.

\paragraph*{\analyst Alone}
Without a \judge, the balanced accuracy of an \analyst ranges from 55.6\%
(Claude \opus) to 66.7\% (\qwen), so even the best model is only 17 points
above a constant answer.
The per-group recall shows where the \analyst fails.
Cases with a vulnerable but not exploitable library (\dg{2}) are handled
reasonably well, with a recall between 66.0\% and 96.3\%.
Exploitable cases (\dg{1}) are harder, with a recall between 35.2\% and 74.1\%;
this result suggests that even frontier models still struggle to perform
multiple-project reasoning.
Cases with a library without known vulnerability (\dg{3}) are the weakest: six of
the eight \analyst models stay below 34\%, and \luna reaches only 9.3\%.
\gemma is the exception at 75.9\%, but it also has the lowest \dg{1} recall
(35.2\%), as it tends to answer \emph{not affected} regardless of the case.
The \analyst alone is therefore not sufficient: it fails to conclude
\emph{not affected} precisely where no known vulnerability anchors its
search, and where a verdict needs the most scrutiny.

\paragraph*{\judge Impact}
In general, pairing \judge and \analyst helps boost software supply-chain
analysis.
The largest gain is observed for \qwen with a four-judge panel, which rises by
15.3 points, from 66.7\% to 81.9\%.
However, the effect is not uniform: 13 of the 18 configurations improve and five
decrease, by up to 3.0 points, and seven of the 13 gains are below five points.
The gains concentrate on a few pairings, such as Claude \sonnet judged by
itself (+16.1), \qwen judged by \luna (+11.6), and the panel for
\gemma and \qwen (+6.9 and +15.3).

A closer look at the results shows that the \judge configuration affects the
improvement more than the choice of \judge model alone.
Specifically, a panel of \judge agents gives the largest and most consistent
gains: both \analyst models gain with a panel, whereas single \judge models
range from $-3.0$ to $+16.1$ points and lower the balanced accuracy in five of
16 pairings.
Among single \judge models, capability does not order the results.
Frontier self-judges (Claude \sonnet, +16.1; Claude \opus, +7.5) and the
cheap \luna on \qwen (+11.6) help, whereas \terra judged by
itself ($-3.0$) and Claude \sonnet judged by Luna ($-3.0$) do not.
The comparison is partly confounded, since most frontier \judge models only
review their own \analyst.

The gains also change the per-group profile.
Claude \sonnet judged by itself raises its \dg{3} recall from 25.9\% to 84.9\%,
and \gemma with the panel from 75.9\% to 92.6\%, while \qwen with Luna or
the panel raises its \dg{1} recall by 22.2 points.
Some gains come at a cost, as Claude \haiku judged by itself loses 11.1
points of \dg{2} recall.
The \judge's ability to tell correct from incorrect \analyst runs does not fully
explain the gains (\autoref{tab:judge-auroc}).
Although Claude \opus and Claude \sonnet have the highest AUROC (0.80 and 0.78)
and can improve their own \analyst, the largest panel gain for \qwen comes
from \judge models with an AUROC between 0.57 and 0.67.
This suggests that part of the benefit comes from the feedback prompting the
\analyst to re-examine its verdict, not only from the \judge identifying errors.
In summary, \judge feedback helps for specific pairings but is not a reliable
fix for the weakness of the \analyst alone, which is why the \judge's report
must also be reliable to be audited by a human (\autoref{sec:trust}).

\finding{Without a \judge, \analyst agents reach only 55.6--66.7\% balanced
accuracy and fail mostly on libraries without a known vulnerability (\dg{3}
recall as low as 9.3\%). Adding a \judge helps in general (13 of 18
configurations) and reaches up to 81.9\% with a four-judge panel.}

\subsection{Judge Reliability (RQ2)}
\label{sec:trust}

The \judge{}'s objective is to determine whether the claims made by the
\analyst are correct.
A human analyst can rely on its report only if the \judge is reliable itself, so
we study whether the claim checks of the \judge are correct and whether its
scores indicate when the verdict of the \analyst is correct.

\paragraph*{Claim Checks}
As the number of claims is too large for a manual review, we review a subset:
for each \analyst model, we randomly select one of the three runs and sample
four cases from each dataset group (\autoref{sec:dataset}), which yields 12
cases per \analyst model and 2,757 of the 12,442 claims (22.2\%).
We check every claim in the \judge report against the recorded evidence to
identify incorrect behaviors in the \judge's checks (details in
\autoref{sec:agentbehavior}).
\autoref{tab:judge-correctness} shows that 85.0\% of the reviewed claims are
correct, and the share grows with the capability of the \judge, from 79.4\% for
\gemma to 93.9\% for Claude \opus.
Claude \opus is clearly better than \luna and Claude \haiku, the \judge models
with the most reviewed claims.
For the other models, fewer than 125 reviewed claims leave wide intervals, so
their ranking is uncertain.

\begin{table}[t]
    \centering
    \captionsetup{width=\columnwidth}
    \caption{Correct claims by \judge model with 95\% Wilson confidence
    intervals and their share among all claims for that \judge in the selected
    runs.}
    \label{tab:judge-correctness}
    \footnotesize
    \begin{tabular}{lrrr}
        \toprule
        \judge model & Reviewed & Correct & 95\% CI \\
        \midrule
        \gemma & 97 (21.4\%) & 77 (79.4\%) & [70.3, 86.2] \\
        \rowcolor{gray!10} \qwen & 445 (22.3\%) & 390 (87.6\%) & [84.3, 90.4] \\
        \muse & 171 (21.3\%) & 145 (84.8\%) & [78.7, 89.4] \\
        \rowcolor{gray!10} \luna & 662 (22.4\%) & 559 (84.4\%) & [81.5, 87.0] \\
        \terra & 77 (20.2\%) & 63 (81.8\%) & [71.8, 88.8] \\
        \rowcolor{gray!10} \sol & 119 (21.9\%) & 100 (84.0\%) & [76.4, 89.5] \\
        Claude \haiku & 886 (22.7\%) & 735 (83.0\%) & [80.3, 85.3] \\
        \rowcolor{gray!10} Claude \sonnet & 121 (21.2\%) & 106 (87.6\%) & [80.6, 92.3] \\
        Claude \opus & 179 (21.6\%) & 168 (93.9\%) & [89.3, 96.5] \\
        \midrule
        Total & 2,757 (22.2\%) & 2,343 (85.0\%) & [83.6, 86.3] \\
        \bottomrule
    \end{tabular}
\end{table}

\paragraph*{Scores and Correctness}
Besides the claim checks, each \judge assigns a total score between 0 and 20 to
the assessment, the sum of ten dimension scores between 0 and 2
(\autoref{design:judge-report}).
A useful score is higher when the verdict of the \analyst is correct.
\autoref{tab:judge-auroc} measures this with the AUROC, the probability that a
randomly chosen correct assessment receives a higher score than a randomly
chosen incorrect one (0.5 is chance), and with the Spearman correlation $\rho$.
Only the \judge models Claude \opus and Claude \sonnet separate correct from
incorrect assessments well (AUROC 0.80 and 0.78, $\rho$ around $+0.5$), whereas
\luna and \terra are close to chance.
The score is therefore a moderate signal: it cannot decide the correctness of a
single assessment, but it helps a human analyst prioritize which assessments to
audit, and only for the strongest \judge models.

\begin{table}[t]
    \centering
    \captionsetup{width=\columnwidth}
    \caption{How well the \judge total score tracks \analyst correctness: AUROC and Spearman $\rho$ with 95\% confidence intervals. Judges are sorted by AUROC. Only the $\rho$ interval of \terra includes zero.}
    \label{tab:judge-auroc}
    \footnotesize
    \setlength{\tabcolsep}{3pt}
    \renewcommand{\arraystretch}{1.1}
    \begin{tabular}{@{}l ll ll@{}}
        \toprule
        & \multicolumn{2}{c}{\textbf{AUROC}} & \multicolumn{2}{c}{\textbf{Spearman $\rho$}} \\
        \cmidrule(lr){2-3}\cmidrule(lr){4-5}
        \textbf{\judge} & value & 95\% CI & value & 95\% CI \\
        \midrule
        Claude \opus & \textbf{0.798} & [0.73, 0.87] & +0.527 & [+0.41, +0.64] \\
        Claude \sonnet & 0.780 & [0.70, 0.87] & +0.487 & [+0.35, +0.62] \\
        \sol & 0.678 & [0.53, 0.82] & +0.321 & [+0.05, +0.56] \\
        \qwen & 0.665 & [0.60, 0.73] & +0.281 & [+0.18, +0.38] \\
        \muse & 0.654 & [0.55, 0.76] & +0.267 & [+0.09, +0.44] \\
        Claude \haiku & 0.643 & [0.57, 0.71] & +0.248 & [+0.14, +0.37] \\
        \gemma & 0.637 & [0.57, 0.70] & +0.265 & [+0.13, +0.40] \\
        \luna & 0.572 & [0.51, 0.63] & +0.127 & [+0.02, +0.23] \\
        \terra & 0.532 & [0.44, 0.62] & +0.058 & [$-$0.11, +0.23] \\
        \bottomrule
    \end{tabular}
\end{table}

\paragraph*{Dimensions}
\autoref{tab:judge-dimension-correlation} breaks the correlation down by
dimension.
The signal comes from the dimensions that check the reasoning behind the
verdict: reachability and vulnerability correlate positively for all nine
\judge models, and classification is the strongest dimension for Claude \sonnet
and Claude \opus ($\rho = +0.60$ and $+0.57$).
In contrast, the dimensions that check process elements, such as scope,
dependency, integration, PoC, tool use, and artifacts, are close to zero.
The \judge models with the best scores are not the ones that yield the largest
gain of the \analyst (\autoref{sec:perform}), so the score helps to audit an
assessment but does not predict how much the feedback improves it.

\begin{table}[t]
    \centering
    \captionsetup{width=\columnwidth}
    \caption{Spearman $\rho$ between each \judge dimension score and \analyst correctness. Columns are \judge models (sorted as in \autoref{tab:judge-auroc}) and their mean, rows are dimensions sorted by the mean. Green marks positive and red negative correlations, with darker shading for larger magnitude.}
    \label{tab:judge-dimension-correlation}
    \scriptsize
    \setlength{\tabcolsep}{0.7pt}
    \renewcommand{\arraystretch}{1.1}
    \begin{tabular}{@{}l *{9}{r} @{\hspace{2pt}} r@{}}
        \toprule
        \textbf{Dimension} & \rotatebox{60}{\textbf{Opus}} & \rotatebox{60}{\textbf{Sonnet}} & \rotatebox{60}{\textbf{Sol}} & \rotatebox{60}{\textbf{Qwen}} & \rotatebox{60}{\textbf{Muse}} & \rotatebox{60}{\textbf{Haiku}} & \rotatebox{60}{\textbf{Gemma}} & \rotatebox{60}{\textbf{Luna}} & \rotatebox{60}{\textbf{Terra}} & \rotatebox{60}{\textit{Mean}} \\
        \midrule
        Reachability & \cellcolor{deltapos!44} 0.41 & \cellcolor{deltapos!29} 0.27 & \cellcolor{deltapos!48} 0.44 & \cellcolor{deltapos!32} 0.30 & \cellcolor{deltapos!35} 0.32 & \cellcolor{deltapos!34} 0.31 & \cellcolor{deltapos!22} 0.20 & \cellcolor{deltapos!32} 0.30 & \cellcolor{deltapos!27} 0.25 & \cellcolor{deltapos!34} \textit{0.31} \\
        Classification & \cellcolor{deltapos!62} 0.57 & \cellcolor{deltapos!65} 0.60 & \cellcolor{deltapos!28} 0.26 & \cellcolor{deltapos!47} 0.43 & \cellcolor{deltapos!25} 0.23 & \cellcolor{deltapos!46} 0.42 & \cellcolor{deltapos!27} 0.25 & \cellcolor{deltapos!6} 0.06 & \cellcolor{deltaneg!11} $-$0.10 & \cellcolor{deltapos!32} \textit{0.30} \\
        Vulnerability & \cellcolor{deltapos!38} 0.35 & \cellcolor{deltapos!43} 0.40 & \cellcolor{deltapos!30} 0.28 & \cellcolor{deltapos!14} 0.13 & \cellcolor{deltapos!27} 0.25 & \cellcolor{deltapos!18} 0.17 & \cellcolor{deltapos!23} 0.21 & \cellcolor{deltapos!25} 0.23 & \cellcolor{deltapos!12} 0.11 & \cellcolor{deltapos!26} \textit{0.24} \\
        Evidence & \cellcolor{deltapos!17} 0.16 & \cellcolor{deltapos!15} 0.14 & \cellcolor{deltapos!36} 0.33 & 0.02 & \cellcolor{deltapos!8} 0.07 & \cellcolor{deltapos!13} 0.12 & \cellcolor{deltapos!21} 0.19 & \cellcolor{deltapos!6} 0.06 & 0.00 & \cellcolor{deltapos!13} \textit{0.12} \\
        Tool Use & \cellcolor{deltapos!21} 0.19 & \cellcolor{deltapos!25} 0.23 & \cellcolor{deltapos!23} 0.21 & $-$0.01 & \cellcolor{deltapos!29} 0.27 & \cellcolor{deltaneg!4} $-$0.04 & \cellcolor{deltapos!10} 0.09 & \cellcolor{deltaneg!5} $-$0.05 & 0.01 & \cellcolor{deltapos!11} \textit{0.10} \\
        Artifacts & \cellcolor{deltapos!4} 0.04 & \cellcolor{deltapos!31} 0.29 & \cellcolor{deltapos!17} 0.16 & \cellcolor{deltapos!13} 0.12 & \cellcolor{deltapos!35} 0.32 & \cellcolor{deltapos!10} 0.09 & \cellcolor{deltaneg!4} $-$0.04 & \cellcolor{deltaneg!11} $-$0.10 & \cellcolor{deltaneg!5} $-$0.05 & \cellcolor{deltapos!10} \textit{0.09} \\
        Scope & \cellcolor{deltapos!16} 0.15 & \cellcolor{deltapos!13} 0.12 & \cellcolor{deltapos!12} 0.11 & 0.00 & \cellcolor{deltapos!5} 0.05 & 0.02 & \cellcolor{deltapos!5} 0.05 & \cellcolor{deltapos!4} 0.04 & \cellcolor{deltaneg!14} $-$0.13 & \cellcolor{deltapos!5} \textit{0.05} \\
        Integration & \cellcolor{deltaneg!12} $-$0.11 & 0.03 & \cellcolor{deltapos!21} 0.19 & \cellcolor{deltapos!14} 0.13 & \cellcolor{deltapos!11} 0.10 & \cellcolor{deltapos!8} 0.07 & \cellcolor{deltapos!4} 0.04 & \cellcolor{deltaneg!9} $-$0.08 & 0.03 & \cellcolor{deltapos!5} \textit{0.05} \\
        PoC & \cellcolor{deltaneg!8} $-$0.07 & \cellcolor{deltapos!9} 0.08 & \cellcolor{deltapos!5} 0.05 & \cellcolor{deltapos!21} 0.19 & \cellcolor{deltapos!12} 0.11 & \cellcolor{deltaneg!4} $-$0.04 & \cellcolor{deltapos!17} 0.16 & \cellcolor{deltaneg!9} $-$0.08 & \cellcolor{deltaneg!31} $-$0.29 & \textit{0.01} \\
        Dependency & \cellcolor{deltapos!6} 0.06 & \cellcolor{deltaneg!8} $-$0.07 & \cellcolor{deltapos!5} 0.05 & 0.01 & \cellcolor{deltaneg!16} $-$0.15 & $-$0.01 & \cellcolor{deltapos!5} 0.05 & \cellcolor{deltaneg!4} $-$0.04 & \cellcolor{deltaneg!12} $-$0.11 & \textit{$-$0.02} \\
        \bottomrule
    \end{tabular}
\end{table}

\finding{The claim checks of the \judge are correct in 85.0\% of the reviewed
claims, up to 93.9\% for Claude \opus. The \judge score is a moderate signal of
the correctness of the \analyst (AUROC 0.53--0.80), driven by reachability,
vulnerability, and classification, and reliable only for the strongest \judge
models (Claude \opus and Claude \sonnet).}

\subsection{Comparison to VEX-Bench (RQ3)}
\label{sec:vexcomp}

VEX-Bench~\cite{vex-bench} is the closest work to \sys on software supply-chain
analysis, as it pursues a similar goal.
We compare the pipelines of \sys and VEX-Bench on our dataset
(\autoref{sec:dataset}), using two open-weight models, \gemma and \qwen, paired
with Codex, with three runs per case.

\paragraph*{Structural Differences}
The main difference between VEX-Bench and \sys lies in their design and the
properties they provide.
VEX-Bench centers around a single agent that performs the analysis and produces
a binary verdict (\emph{affected} or \emph{not affected}), a predicted category,
and a short justification, which resembles the purpose of the \judge's report.
However, VEX-Bench does not include any mechanism to produce auditable
information about the analysis process, which is the main goal of \sys.
Moreover, VEX-Bench is not designed to create PoCs, which are strong evidence of
exploitability when they trigger the vulnerability.
Conversely, \sys provides auditable and verifiable claims that (i) improve the
performance of the \analyst and (ii) human analysts can rely on during their
security vetting.
\autoref{tab:vex-overview} compares further properties of both pipelines.
In exchange, VEX-Bench is much cheaper than \sys: it uses about ten times fewer
tokens with \gemma and 80 times fewer with \qwen (Appendix~\ref{app:cost}).

\paragraph*{Performance Comparison}
\autoref{tab:vexresult} lists the results of VEX-Bench on our dataset.
The balanced accuracy of VEX-Bench is comparable to \sys without a \judge, but lower
than that of \sys with a \judge in most configurations, which indicates the
importance of the \judge.
For \gemma, only the \judge panel performs better than VEX-Bench, and this gain
is small.
For \qwen, \sys with a \judge is higher in all three \judge configurations, and
by 12 percentage points with the \judge panel.
In addition, the recall of VEX-Bench on cases with an exploitable library
vulnerability (\dg{1}) is much lower than on cases without an exploitable
vulnerability (\dg{2} and \dg{3}).
Notably, \qwen reaches perfect recall on the cases with a library vulnerability
that is not exploitable (\dg{2}).
Together with the low recall on \dg{1}, this shows that VEX-Bench is heavily
biased towards the label \emph{not affected}.
A possible explanation is that the agent has no access to the library source
code.

\finding{VEX-Bench performs similarly to \sys without a \judge, but it is biased
towards \emph{not affected}, with a low recall on exploitable cases (\dg{1}).
Combining the \analyst with a \judge yields higher performance, and the
independent \judge provides auditable logs that VEX-Bench lacks.}

\begin{table}[t]
    \centering
    \captionsetup{width=\columnwidth}
    \caption{Overview of VEX-Bench and \sys on our dataset. Balanced accuracy of \sys + \judge is the best configuration (panel). Bal.\ acc.: balanced accuracy. Tokens: median per case (Appendix~\ref{app:cost}); the \judge tokens were not recorded, so \sys + \judge is a lower bound. Bold marks the best value in each row.}
    \label{tab:vex-overview}
    \newcommand{\cfull}{\tikz[baseline=-0.6ex]\fill (0,0) circle (0.6ex);}
    \newcommand{\cempty}{\tikz[baseline=-0.6ex]\draw (0,0) circle (0.6ex);}
    \newcommand{\chalf}{\tikz[baseline=-0.6ex]{\draw (0,0) circle (0.6ex); \fill (0,-0.6ex) arc (-90:90:0.6ex) -- cycle;}}
    \footnotesize
    \setlength{\tabcolsep}{4pt}
    \renewcommand{\arraystretch}{1.1}
    \begin{tabular}{@{}l ccc@{}}
        \toprule
        & \textbf{VEX-Bench} & \textbf{\sys} & \textbf{\sys+\judge} \\
        \midrule
        Library source & \cempty & \cfull & \cfull \\
        \rowcolor{gray!12} Structured claims & \cempty & \cfull & \cfull \\
        PoC artifact & \cempty & \cfull & \cfull \\
        \rowcolor{gray!12} Auditable & \cempty & \chalf & \cfull \\
        Independent \judge & \cempty & \cempty & \cfull \\
        \midrule
        \rowcolor{gray!12} Bal.\ acc., \gemma & 0.644 & 0.583 & \textbf{0.653} \\
        Bal.\ acc., \qwen & 0.699 & 0.667 & \textbf{0.819} \\
        \rowcolor{gray!12} Tokens, \gemma & \textbf{170k} & 1.70M & $>$1.70M \\
        Tokens, \qwen & \textbf{90.5k} & 7.61M & $>$7.61M \\
        \bottomrule
    \end{tabular}
    \par\smallskip
    \noindent\begin{minipage}{\columnwidth}\footnotesize\raggedright
        \cfull~yes, \chalf~partial (the trace is recorded, but the claims are not checked against it), \cempty~no. \emph{Auditable}: claims linked to a tool-call trace and checked against it. \emph{PoC artifact}: produced for \emph{affected} verdicts.
    \end{minipage}
\end{table}

\begin{table}[t]
    \centering
    \captionsetup{width=\columnwidth}
    \caption{VEX-Bench results on our dataset (mean $\pm$ SD over runs): balanced accuracy and recall per dataset group. Bold marks the best value in each column. \dg{1}: exploitable library vulnerability; \dg{2}: library vulnerability that is not exploitable; \dg{3}: library without known vulnerability.}
    \label{tab:vexresult}
    \footnotesize
    \setlength{\tabcolsep}{4pt}
    \renewcommand{\arraystretch}{1.1}
    \begin{tabular}{@{}l r ccc@{}}
        \toprule
        & \multicolumn{1}{c}{\textbf{Balanced}} & \multicolumn{3}{c}{\textbf{Recall by dataset group}} \\
        \cmidrule(lr){3-5}
        \textbf{\analyst} & \multicolumn{1}{c}{\textbf{accuracy}} & \dg{1} & \dg{2} & \dg{3} \\
        \midrule
        Gemma 4 & 0.644{\scriptsize\,$\pm$\,0.021} & 0.315{\scriptsize\,$\pm$\,0.032} & 0.981{\scriptsize\,$\pm$\,0.032} & 0.963{\scriptsize\,$\pm$\,0.064} \\
        \rowcolor{gray!12} Qwen 3.8 & \textbf{0.699}{\scriptsize\,$\pm$\,0.045} & \textbf{0.407}{\scriptsize\,$\pm$\,0.085} & \textbf{1.000}{\scriptsize\,$\pm$\,0.000} & \textbf{0.981}{\scriptsize\,$\pm$\,0.032} \\
        \bottomrule
    \end{tabular}
\end{table}

\subsection{Agents' Behavior (RQ4)}
\label{sec:agentbehavior}

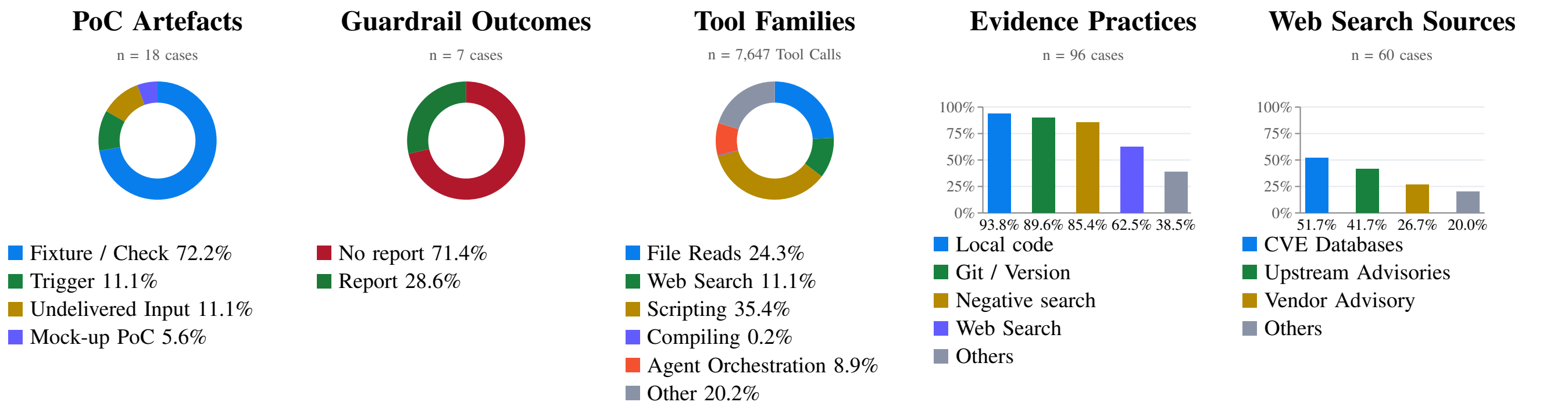
\begin{figure*}[!t]
    \centering
    \resizebox{\textwidth}{!}{\begin{tikzpicture}[
  font=\small,
  every node/.style={inner sep=0pt}
]
\definecolor{chartTrack}{HTML}{E7EBF0}
\definecolor{chartBlue}{HTML}{087EED}
\definecolor{chartTeal}{HTML}{18813D}
\definecolor{chartGold}{HTML}{B58900}
\definecolor{chartViolet}{HTML}{625CFF}
\definecolor{chartOrange}{HTML}{F35130}
\definecolor{chartGray}{HTML}{8A93A6}
\definecolor{chartBlueLight}{HTML}{5F91B3}
\definecolor{chartTealLight}{HTML}{58A197}
\definecolor{chartGoldLight}{HTML}{D7A644}
\definecolor{chartVioletLight}{HTML}{9A8CBD}
\definecolor{chartOlive}{HTML}{82914D}
\definecolor{chartGreen}{HTML}{5C8050}

\newcommand{\scarvaShareRing}[5]{%
  \begin{scope}[shift={(#1,#2)}]
    \draw[chartTrack,line width=2.4pt] (0,0) circle[radius=#3];
    \draw[#5,line width=2.4pt,line cap=round]
      (90:#3) arc[start angle=90,end angle=#4,radius=#3];
  \end{scope}%
}
\newcommand{\scarvaPieSlice}[7]{%
  \begin{scope}[shift={(#1,#2)}]
    \path[fill=#7,draw=none]
      (#5:#3) arc[start angle=#5,end angle=#6,radius=#3]
      -- (#6:#4) arc[start angle=#6,end angle=#5,radius=#4]
      -- cycle;
  \end{scope}%
}
\newcommand{\scarvaBar}[4]{%
  \begin{scope}[shift={(#1,#2)}]
    \path[fill=#4,draw=none] (-0.13,0) rectangle (0.13,#3);
  \end{scope}%
}

\path[use as bounding box] (-8.8,0.25) rectangle (8.8,4.9);

\node[align=center,text width=3.25cm,font=\small\bfseries] at (-7,4.56)
  {PoC Artefacts};
\node[align=center,font=\tiny,text=black!65] at (-7,4.17)
  {n = 18 cases};
\node[align=center,text width=3.25cm,font=\small\bfseries] at (-3.5,4.56)
  {Guardrail Outcomes};
\node[align=center,font=\tiny,text=black!65] at (-3.5,4.17)
  {n = 7 cases};
\node[align=center,text width=3.25cm,font=\small\bfseries] at (0,4.56)
  {Tool Families};
\node[align=center,font=\tiny,text=black!65] at (0,4.17)
  {n = 7,647 Tool Calls};
\node[align=center,text width=3.25cm,font=\small\bfseries] at (3.5,4.56)
  {Evidence Practices};
\node[align=center,font=\tiny,text=black!65] at (3.5,4.17)
  {n = 96 cases};
\node[align=center,text width=3.25cm,font=\small\bfseries] at (7,4.56)
  {Web Search Sources};
\node[align=center,font=\tiny,text=black!65] at (7,4.17)
  {n = 60 cases};

\scarvaPieSlice{-7}{3.2}{0.67}{0.43}{90}{-170}{chartBlue}
\scarvaPieSlice{-7}{3.2}{0.67}{0.43}{-170}{-210}{chartTeal}
\scarvaPieSlice{-7}{3.2}{0.67}{0.43}{-210}{-250}{chartGold}
\scarvaPieSlice{-7}{3.2}{0.67}{0.43}{-250}{-270}{chartViolet}
\node[align=left,anchor=north west,text width=3.4cm,font=\scriptsize] at (-8.7,2.02) {%
  \textcolor{chartBlue}{\rule{0.65em}{0.65em}}\hspace{0.35em}Fixture / Check 72.2\%\\[1pt]
  \textcolor{chartTeal}{\rule{0.65em}{0.65em}}\hspace{0.35em}Trigger 11.1\%\\[1pt]
  \textcolor{chartGold}{\rule{0.65em}{0.65em}}\hspace{0.35em}Undelivered Input 11.1\%\\[1pt]
  \textcolor{chartViolet}{\rule{0.65em}{0.65em}}\hspace{0.35em}Mock-up PoC 5.6\%
};

\scarvaPieSlice{-3.5}{3.2}{0.67}{0.43}{90}{-167.143}{deltaneg}
\scarvaPieSlice{-3.5}{3.2}{0.67}{0.43}{-167.143}{-270}{deltapos}
\node[align=left,anchor=north west,text width=3.4cm,font=\scriptsize] at (-5.2,2.02) {%
  \textcolor{deltaneg}{\rule{0.65em}{0.65em}}\hspace{0.35em}No report 71.4\%\\[1pt]
  \textcolor{deltapos}{\rule{0.65em}{0.65em}}\hspace{0.35em}Report 28.6\%
};

\scarvaPieSlice{0}{3.2}{0.67}{0.43}{90}{2.625}{chartBlue}
\scarvaPieSlice{0}{3.2}{0.67}{0.43}{2.625}{-37.344}{chartTeal}
\scarvaPieSlice{0}{3.2}{0.67}{0.43}{-37.344}{-164.735}{chartGold}
\scarvaPieSlice{0}{3.2}{0.67}{0.43}{-164.735}{-165.488}{chartViolet}
\scarvaPieSlice{0}{3.2}{0.67}{0.43}{-165.488}{-197.360}{chartOrange}
\scarvaPieSlice{0}{3.2}{0.67}{0.43}{-197.360}{-270}{chartGray}
\node[align=left,anchor=north west,text width=3.4cm,font=\scriptsize] at (-1.7,2.02) {%
  \textcolor{chartBlue}{\rule{0.65em}{0.65em}}\hspace{0.35em}File Reads 24.3\%\\[1pt]
  \textcolor{chartTeal}{\rule{0.65em}{0.65em}}\hspace{0.35em}Web Search 11.1\%\\[1pt]
  \textcolor{chartGold}{\rule{0.65em}{0.65em}}\hspace{0.35em}Scripting 35.4\%\\[1pt]
  \textcolor{chartViolet}{\rule{0.65em}{0.65em}}\hspace{0.35em}Compiling 0.2\%\\[1pt]
  \textcolor{chartOrange}{\rule{0.65em}{0.65em}}\hspace{0.35em}Agent Orchestration 8.9\%\\[1pt]
  \textcolor{chartGray}{\rule{0.65em}{0.65em}}\hspace{0.35em}Other 20.2\%
};

\draw[chartTrack,line width=0.4pt] (2.36,2.68) -- (4.72,2.68);
\draw[chartTrack,line width=0.4pt] (2.36,2.98) -- (4.72,2.98);
\draw[chartTrack,line width=0.4pt] (2.36,3.28) -- (4.72,3.28);
\draw[chartTrack,line width=0.4pt] (2.36,3.58) -- (4.72,3.58);
\draw[black!45,line width=0.4pt] (2.36,2.38) -- (4.72,2.38);
\draw[black!45,line width=0.4pt] (2.36,2.38) -- (2.36,3.58);
\foreach \y in {2.38,2.68,2.98,3.28,3.58} {
  \draw[black!45,line width=0.4pt] (2.30,\y) -- (2.36,\y);
}
\node[font=\tiny,text=black!65,anchor=east] at (2.28,2.38) {0\%};
\node[font=\tiny,text=black!65,anchor=east] at (2.28,2.68) {25\%};
\node[font=\tiny,text=black!65,anchor=east] at (2.28,2.98) {50\%};
\node[font=\tiny,text=black!65,anchor=east] at (2.28,3.28) {75\%};
\node[font=\tiny,text=black!65,anchor=east] at (2.28,3.58) {100\%};
\scarvaBar{2.55}{2.38}{1.126}{chartBlue}
\scarvaBar{3.05}{2.38}{1.075}{chartTeal}
\scarvaBar{3.55}{2.38}{1.025}{chartGold}
\scarvaBar{4.05}{2.38}{0.750}{chartViolet}
\scarvaBar{4.55}{2.38}{0.462}{chartGray}
\node[font=\tiny] at (2.55,2.25) {93.8\%};
\node[font=\tiny] at (3.05,2.25) {89.6\%};
\node[font=\tiny] at (3.55,2.25) {85.4\%};
\node[font=\tiny] at (4.05,2.25) {62.5\%};
\node[font=\tiny] at (4.55,2.25) {38.5\%};
\node[align=left,anchor=north west,text width=3.4cm,font=\scriptsize] at (1.8,2.12) {%
  \textcolor{chartBlue}{\rule{0.65em}{0.65em}}\hspace{0.35em}Local code\\[1pt]
  \textcolor{chartTeal}{\rule{0.65em}{0.65em}}\hspace{0.35em}Git / Version\\[1pt]
  \textcolor{chartGold}{\rule{0.65em}{0.65em}}\hspace{0.35em}Negative search\\[1pt]
  \textcolor{chartViolet}{\rule{0.65em}{0.65em}}\hspace{0.35em}Web Search\\[1pt]
  \textcolor{chartGray}{\rule{0.65em}{0.65em}}\hspace{0.35em}Others
};

\draw[chartTrack,line width=0.4pt] (5.96,2.68) -- (8.05,2.68);
\draw[chartTrack,line width=0.4pt] (5.96,2.98) -- (8.05,2.98);
\draw[chartTrack,line width=0.4pt] (5.96,3.28) -- (8.05,3.28);
\draw[chartTrack,line width=0.4pt] (5.96,3.58) -- (8.05,3.58);
\draw[black!45,line width=0.4pt] (5.96,2.38) -- (8.05,2.38);
\draw[black!45,line width=0.4pt] (5.96,2.38) -- (5.96,3.58);
\foreach \y in {2.38,2.68,2.98,3.28,3.58} {
  \draw[black!45,line width=0.4pt] (5.90,\y) -- (5.96,\y);
}
\node[font=\tiny,text=black!65,anchor=east] at (5.88,2.38) {0\%};
\node[font=\tiny,text=black!65,anchor=east] at (5.88,2.68) {25\%};
\node[font=\tiny,text=black!65,anchor=east] at (5.88,2.98) {50\%};
\node[font=\tiny,text=black!65,anchor=east] at (5.88,3.28) {75\%};
\node[font=\tiny,text=black!65,anchor=east] at (5.88,3.58) {100\%};
\scarvaBar{6.15}{2.38}{0.620}{chartBlue}
\scarvaBar{6.72}{2.38}{0.500}{chartTeal}
\scarvaBar{7.29}{2.38}{0.320}{chartGold}
\scarvaBar{7.86}{2.38}{0.240}{chartGray}
\node[font=\tiny] at (6.15,2.25) {51.7\%};
\node[font=\tiny] at (6.72,2.25) {41.7\%};
\node[font=\tiny] at (7.29,2.25) {26.7\%};
\node[font=\tiny] at (7.86,2.25) {20.0\%};
\node[align=left,anchor=north west,text width=3.4cm,font=\scriptsize] at (5.3,2.12) {%
  \textcolor{chartBlue}{\rule{0.65em}{0.65em}}\hspace{0.35em}CVE Databases\\[1pt]
  \textcolor{chartTeal}{\rule{0.65em}{0.65em}}\hspace{0.35em}Upstream Advisories\\[1pt]
  \textcolor{chartGold}{\rule{0.65em}{0.65em}}\hspace{0.35em}Vendor Advisory\\[1pt]
  \textcolor{chartGray}{\rule{0.65em}{0.65em}}\hspace{0.35em}Others
};
\end{tikzpicture}}
    \caption{%
        \sys \analyst behavior across the 96 manually reviewed cases from 8 runs (12 cases per run, 4 per dataset group).
    }
    \label{fig:analyst-behavior-pies}
\end{figure*}

This section analyzes how the \analyst{s} solve the software supply-chain problem and reveals noteworthy behaviors.
The findings highlight the most common agent misbehaviors and can guide future post-training efforts.
For our analysis, we consider the manually selected claims discussed in
\autoref{sec:trust}.
All reported numbers refer to \autoref{fig:analyst-behavior-pies} and
\autoref{tab:tool-usage}.

\paragraph*{Evidence and Applicability}
The \analyst{s} make extensive use of git commands to identify vulnerable code
(86 \texttt{git log}).
In 37 assessments, the \analyst treats missing evidence as evidence, although
missing evidence does not establish that the vulnerable behavior is absent.

The \analyst also consults external sources to collect CVE information and support
the analysis.
The claims cite at least one external source in 60 assessments (62.5\%),
including CVE databases in 31 (51.7\%), upstream advisories in 25 (41.7\%),
vendor advisories in 16 (26.7\%), and other external references in 12 (20.0\%); source types overlap.
These numbers describe what the \analyst reports as evidence. The numbers do not
establish that every cited source was successfully retrieved.
Out of 11 claims asserting that external material was obtained, the \judge{s}
mark four as supported, four as partially supported, one as unsupported, and two
as contradicted.
In the two contradicted claims, the \judge finds no recorded retrieval of the
claimed advisory.
Hence, the \analyst either hallucinates the content or recalls it from its
training data.

In nine of the 32 assessments with ground-truth label \emph{affected}, the
\analyst uses a newer fixed source instead of the vulnerable version, and
concludes that the host application is \emph{not affected}.
For example, Gemma 4 assesses an OpenSSL 4.0 checkout as fixed, while the
dataset specifies the vulnerable 1.1.1m release.
All four judges rate this classification as correct with concerns.
Across the nine assessments, all 28 \judge reports rate the classifications as
correct, although at least two warn that the conclusion could overgeneralize
to historical versions.
The repository setup and prompt do not enforce the dataset's version
constraints, so the \analyst can take shortcuts and analyze a newer revision
(see \autoref{sec:discussion}).

In 21 assessments, the \analyst produces an incomplete analysis and reports
missing evidence or open questions.
In other cases, the \analyst accepts an incomplete exploitability chain as valid.
Among the 14 assessments marked as complete chains, 11 contain at least one
link marked as \emph{likely} rather than \emph{proven}.
Appendix~\ref{app:prompts} contains the system prompts used.

\paragraph*{Proof of Concept}
The \analyst creates a PoC in all 18 assessments classified as \emph{affected}.
One PoC is a mock, and two implement a trigger for the host application: one
is executed and reproduces the vulnerable behavior, while the other is not
executed.
Thirteen artifacts implement only local fixtures or checks, and the other two
construct attack inputs without delivering them to the host application.

\gemma produces one of its two PoC artifacts as a mock that prints a
claimed outcome without performing any exploit.
Three out of the four \judge agents flag this behavior correctly, and
only \muse accepts the mock as valid.
OpenAI's models produce checks or simulations rather than an executable exploit,
and \opus produces an attack-input template and a benign reachability
probe, but none obtains a complete trigger.
Conversely, \qwen produces more working PoCs: three of the analyzed ones
work correctly.
In one assessment, \qwen builds the host application and tests vulnerable and
fixed library versions, producing the only observed host-application trigger.
Other models use fixtures, static checks, or unexecuted scripts and do not run
or build the host application from source.
Six fixtures are executed in total, but they demonstrate only a selected
behavior rather than the complete exploit. 
The small fraction of working PoCs indicates that producing such
artifacts does not suffice to prove a vulnerability, which
motivates the use of a \judge.

\paragraph*{Cybersecurity Guardrails}
Proprietary models have security guardrails that can terminate sessions when
possible misuse is detected.
The Anthropic account is enrolled in the Cyber Verification Program (with more
lenient guardrails), whereas the OpenAI account has no equivalent
verification.
These different account conditions limit the comparison between the providers
(\autoref{sec:discussion}).

A cybersecurity-risk flag appears in seven of the 72 analyzed assessments:
four involve GPT-5.6 Terra and three involve GPT-5.6 Sol.
In all cases, the guardrails trigger early in the analysis and impede
the production of a complete assessment.
One of the three Sol assessments reaches repository inspection and vulnerability
research, while the other two are still establishing the scope and inventory
before the OpenAI servers terminate the connection.
Of the four Terra assessments, one stops during source analysis and one near the
start, while the other two produce a final assessment but are flagged before
creating all artifacts and reports.
The \haiku \judge rejects one and accepts the other with concerns, which shows
how a second \judge aids the validation process.

\paragraph*{Tool Usage}
The \analyst uses tool calls to read source code, search the web, inspect
repository history, compile code, and run tests.
As shown in \autoref{tab:tool-usage}, \qwen uses the most tool calls on average,
and issues 1,330 of its 1,501 recorded calls as shell commands for source
inspection, builds, and tests.
Among the Claude models, \sonnet records only 17.6 tool calls on average.
As \qwen and \sonnet both reach a high balanced accuracy, we find no evidence that
issuing more tool calls correlates with performance.

\qwen produces the most claims, 23.2 per run on average, while \gemma and \terra
produce the fewest.
We observe that the number of claims does not correlate with the number of tool
calls: \sonnet issues the fewest tool calls but records more claims than four
other models.

\begin{table}[t]
    \centering
    \captionsetup{width=\columnwidth}
    \caption{Average tool calls and claims per assessment for each \analyst model. The last column is derived as claims per 10 tool calls. Bold marks the highest value in each column.}
    \label{tab:tool-usage}
    \footnotesize
    \setlength{\tabcolsep}{4pt}
    \renewcommand{\arraystretch}{1.1}
    \begin{tabular}{@{}l r r r@{}}
        \toprule
        \textbf{\analyst} & \textbf{Tool calls} & \textbf{Claims} & \textbf{Claims / 10 calls} \\
        \midrule
        Gemma 4 & 39.2 & 4.2 & 1.07 \\
        Qwen 3.8 & \textbf{125.1} & \textbf{23.2} & 1.85 \\
        \midrule
        GPT-5.6 Luna & 66.2 & 8.3 & 1.25 \\
        GPT-5.6 Terra & 33.0 & 7.6 & 2.30 \\
        GPT-5.6 Sol & 89.8 & 10.7 & 1.19 \\
        \midrule
        Claude Haiku 5.5 & 40.5 & 14.1 & 3.48 \\
        Claude Sonnet 5.5 & 17.6 & 10.9 & \textbf{6.19} \\
        Claude Opus 5.5 & 26.8 & 14.3 & 5.34 \\
        \bottomrule
    \end{tabular}
\end{table}

\finding{Agents produce plausible but weakly grounded evidence: few PoCs reach
the host application, nine affected cases drift to a fixed revision, and some
cited advisories are never retrieved. The \judge and the logs expose these
failures.}

\section{Discussion}
\label{sec:discussion}

\paragraph*{Context Boundaries}
In nine \emph{affected} assessments, the \analyst moves to a newer fixed
revision because neither the repository setup nor the prompt enforces the
version constraints of the case.
This behavior shows a misalignment with the task, and stricter boundaries
would likely remove the shortcut and improve the \analyst's results.

\paragraph*{Limitations}
The dataset contains 54 C/C++ cases, 18 per group, and the reported differences
are not tested for significance at the case level.
Some CVEs may predate the model cutoffs (\autoref{tab:model-cutoffs}), so the
\analyst can recall the answer rather than derive it.
The frontier models are judged only by themselves, which confounds the \judge
comparison.
The Anthropic account runs under the Cyber Verification Program, whereas the
OpenAI account does not, so the guardrail flags limit the comparison between
the two providers.

\section{Conclusion}
\label{sec:conclusion}

This work presents \sys, the first auditable agentic framework for software
supply-chain analysis.
By pairing an \analyst with a \judge agent, \sys produces auditable evidence.
A \judge panel raises the balanced accuracy of \qwen from 0.667 to 0.819, and
the logs let users verify each verdict.
Our evaluation shows that an \analyst alone is not a reliable exploitability
oracle: the \analyst tends to answer \emph{not affected} and rarely recognizes
host applications without a known vulnerability (\dg{3}).
Although a \judge improves the outcome in most pairings, the gain depends on the
pairing rather than on the model tier, and only a \judge panel is
consistently effective.
Even when accuracy is modest, the claims, artifacts, and \judge reports let
users audit the verdict, which a bare label does not offer.

\clearpage

\section*{Research Ethics Considerations}
\label{sec:ethics}

\noindent \textbf{Stakeholders and data.}
Our study involves no human subjects and no personal data, so no IRB review
applies.
The cases consist of publicly disclosed vulnerabilities and open-source
repositories, mounted read-only in the containers.
The stakeholders are the maintainers and users of the analyzed host applications
and libraries, and the providers of the LLM services we query.

\noindent \textbf{Live systems and vulnerabilities.}
The agents analyze source code inside isolated Docker containers.
Any proof of concept runs locally in the container, and we never test a
third-party deployment.
Web access is limited to public sources such as advisories and documentation
at low request volume.
All vulnerabilities in the dataset are public and already patched or documented.
We did not identify previously unknown vulnerabilities.

\noindent \textbf{Dual use and misuse.}
\sys targets defenders who triage whether a known vulnerability affects an
application.
Its \analyst agent can also describe how to trigger a known flaw, which could
lower the effort of exploiting unpatched systems.
We limit this risk by working only with public CVEs by running the agents in a
sandbox.
We judge that the benefit of auditable verdicts, which reduce alert fatigue and
unsupported claims, outweighs this marginal risk, as it helps maintainers to
implement mitigations in a timely manner.
A second harm is over-trust: a wrong \emph{not affected} verdict can leave a
system unpatched.
\sys is therefore decision support for a human analyst, and we report where the
\judge fails to detect errors (\autoref{sec:agentbehavior}).

\noindent \textbf{LLM providers and compute.}
We follow the usage policies of all providers and do not attempt to bypass
their safety mechanisms.
Requests that a provider's guardrails flag are aborted and reported as such.
The Anthropic account is enrolled in the Cyber Verification Program, while the
open-weight models run on a server with four NVIDIA H100 GPUs.

\bibliographystyle{IEEEtran}
\bibliography{references}

\appendices
\section{Agent Prompts}
\label{app:prompts}

\sys uses three prompts, whose role is described in \autoref{impl:runtime}.
Their exact text is available in our anonymized repository.
\begin{itemize}
    \item \analyst prompt: a template that we fill in per case with the
    CVE, the host application, and the library. The repository contains all 54
    instances, \eg case 1:
    \url{https://anonymous.4open.science/r/scarva/docker/dataset/scarva/first_itteration/case-001/GOAL.md}.
    In the second iteration, the prompt adds one paragraph that points to the
    \judge feedback.
    \item \analyst feedback prompt (second iteration only):
    \url{https://anonymous.4open.science/r/scarva/docker/dataset/scarva/second_itteration/JUDGEMENT.md}.
    \item \judge prompt:
    \url{https://anonymous.4open.science/r/scarva/docker/judge/GOAL.md}.
\end{itemize}

\section{Dataset}
\label{app:dataset}

\autoref{tab:scarva-cves} lists the CVEs, the host applications, and the
libraries of the cases in \dg{1} and \dg{2}.
All the CVEs of our dataset are publicly disclosed and were not discovered
during the development of \sys.

\begin{table}[t]
    \centering
    \captionsetup{width=\columnwidth}
    \caption{The CVEs used in the SCARVA dataset.}
    \label{tab:scarva-cves}
    \begin{tabular}{lll}
        \toprule
        CVE ID & Host Application / Library & Description \\
        \midrule
        CVE-2014-0160 & nginx / OpenSSL & Buffer over-read \\
        \rowcolor{gray!10} CVE-2014-0160 & MongoDB Server / OpenSSL & Buffer over-read \\
        CVE-2022-0778 & nginx / OpenSSL & Infinite loop \\
        \rowcolor{gray!10} CVE-2022-0778 & MongoDB Server / OpenSSL & Infinite loop \\
        CVE-2021-3449 & nginx / OpenSSL & Null dereference \\
        \rowcolor{gray!10} CVE-2021-3449 & MongoDB Server / OpenSSL & Null dereference \\
        CVE-2025-14847 & MongoDB Server / zlib & Memory disclosure \\
        \rowcolor{gray!10} CVE-2024-3094 & OpenSSH sshd / xz / liblzma & Backdoor \\
        CVE-2023-45288 & Traefik / Go net/http & DoS \\
        \rowcolor{gray!10} CVE-2024-28182 & nghttpx / libnghttp2 & DoS \\
        CVE-2023-38545 & curl / libcurl & Heap overflow \\
        \rowcolor{gray!10} CVE-2023-4863 & Google Chrome / libwebp & Heap overflow \\
        CVE-2023-4863 & Electron / libwebp & Heap overflow \\
        \rowcolor{gray!10} CVE-2023-4863 & Firefox / libwebp & Heap overflow \\
        CVE-2023-44487 & Apache httpd / libnghttp2 & DoS \\
        \rowcolor{gray!10} CVE-2023-44487 & nginx / httpv2 & DoS \\
        CVE-2024-6387 & OpenSSH sshd / glibc & Race condition \\
        \rowcolor{gray!10} CVE-2024-4577 & Apache httpd / PHP-CGI & Argument injection \\
        CVE-2025-9232 & nginx / OpenSSL & Out-of-bounds read \\
        \rowcolor{gray!10} CVE-2021-33560 & nginx / libgcrypt & Side-channel attack \\
        CVE-2020-36846 & MongoDB Server / Brotli & Buffer overflow \\
        \rowcolor{gray!10} CVE-2023-43787 & Wireshark / libX11 & Integer overflow \\
        CVE-2023-45853 & Apache httpd / zlib & Heap overflow \\
        \rowcolor{gray!10} CVE-2023-45853 & Git / zlib & Heap overflow \\
        CVE-2023-38545 & Apache httpd / libcurl & Heap overflow \\
        \rowcolor{gray!10} CVE-2023-38545 & Git / libcurl & Heap overflow \\
        CVE-2023-0464 & OpenConnect / OpenSSL & DoS \\
        \rowcolor{gray!10} CVE-2023-0215 & curl / OpenSSL & Use-after-free \\
        CVE-2023-0401 & HAProxy / OpenSSL & Null dereference \\
        \rowcolor{gray!10} CVE-2022-37434 & nginx / zlib & Buffer overflow \\
        CVE-2023-4863 & Wireshark / libwebp & Heap overflow \\
        \rowcolor{gray!10} CVE-2022-27404 & Wireshark / FreeType & Heap overflow \\
        CVE-2024-25062 & OpenConnect / libxml2 & Use-after-free \\
        \rowcolor{gray!10} CVE-2024-34459 & qpdf / libxml2 & Buffer over-read \\
        CVE-2024-0727 & rsync / OpenSSL & Null dereference \\
        \rowcolor{gray!10} OSV-2024-636 & MongoDB Server / OpenSSL & Null dereference \\
        \bottomrule
    \end{tabular}
\end{table}

\section{Cost}
\label{app:cost}

For each \analyst model, we compute the median number of tokens, the median
number of output tokens, the median cost in USD, and the median wall-clock time
of one assessment over the 54 cases (\autoref{tab:analyst-cost}).
The cost of the \judge is not included, as its token usage was not recorded.
\qwen consumes the most tokens (7.61M) and Claude \sonnet the least (0.56M).
In USD, \sol is the most expensive model (\$3.95) and \luna the cheapest
commercial one (\$0.09), whereas \gemma and \qwen are self-hosted and have no
price.
The price does not track accuracy: without a \judge, \luna outperforms \sol
(\autoref{tab:results}).

\begin{table}[t]
    \centering
    \captionsetup{width=\columnwidth}
    \caption{Median cost and wall-clock time of one \analyst assessment without \judge over the 54 cases, with 95\% confidence intervals.
    Cost uses list prices of November 2026.}
   \label{tab:analyst-cost}
    \scriptsize
    \setlength{\tabcolsep}{4pt}
    \renewcommand{\arraystretch}{1.1}
    \begin{tabular}{@{}l r r r r@{}}
        \toprule
        \textbf{\analyst} & \textbf{Tokens (M)} & \textbf{Output (k)} & \textbf{Cost (USD)} & \textbf{Time (min)} \\
        \midrule
        Gemma 4 & 1.70 {\tiny[1.46, 1.98]} & 12.9 {\tiny[11.9, 13.4]} & n/a & 11.4 {\tiny[9.8, 13.6]} \\
        \rowcolor{gray!12} Qwen 3.8 & 7.61 {\tiny[6.90, 8.88]} & 121.1 {\tiny[105.3, 131.7]} & n/a & 35.6 {\tiny[33.6, 40.1]} \\
        GPT-5.6 Luna & 1.80 {\tiny[1.58, 2.06]} & 20.4 {\tiny[18.4, 22.4]} & 0.09 {\tiny[0.09, 0.10]} & 4.9 {\tiny[4.7, 5.2]} \\
        \rowcolor{gray!12} GPT-5.6 Terra & 1.57 {\tiny[1.46, 1.74]} & 19.0 {\tiny[18.4, 19.6]} & 0.85 {\tiny[0.77, 0.92]} & 3.7 {\tiny[3.4, 3.9]} \\
        GPT-5.6 Sol & 4.58 {\tiny[3.66, 5.01]} & 39.0 {\tiny[36.6, 43.1]} & 3.95 {\tiny[3.46, 4.30]} & 7.4 {\tiny[6.7, 8.4]} \\
        \rowcolor{gray!12} Claude Haiku 5.5 & 2.17 {\tiny[1.85, 2.54]} & 50.8 {\tiny[46.9, 54.9]} & 0.12 {\tiny[0.08, 0.15]} & 5.3 {\tiny[5.1, 6.0]} \\
        Claude Sonnet 5.5 & 0.56 {\tiny[0.50, 0.62]} & 14.7 {\tiny[13.8, 15.6]} & 0.32 {\tiny[0.30, 0.34]} & 2.2 {\tiny[2.1, 2.4]} \\
        \rowcolor{gray!12} Claude Opus 5.5 & 1.28 {\tiny[1.14, 1.56]} & 27.1 {\tiny[25.2, 30.4]} & 1.14 {\tiny[1.03, 1.33]} & 5.6 {\tiny[4.9, 6.4]} \\
        \bottomrule
    \end{tabular}
\end{table}

Compared with VEX-Bench, the \analyst of \sys uses about ten times more tokens
with \gemma (1.70M against 170k) and about 80 times more with \qwen (7.61M
against 90.5k), which reflects its longer, tool-based investigation of both
repositories.

\section{Alternative Policies}
\label{app:policies}

\autoref{tab:results} reports the results under the \emph{conservative} policy,
which treats \emph{potentially affected} verdicts as \emph{affected}.
This appendix reports the same results under the \emph{relaxed} policy, which
treats \emph{potentially affected} verdicts as \emph{not affected}, and under
the \emph{excluded} policy, which ignores the cases with a \emph{potentially
affected} verdict (\autoref{sec:dataset}).
\begin{table}[t]
    \centering
    \captionsetup{width=\columnwidth}
    \caption{Balanced accuracy under the three policies for \emph{potentially affected} classification by the \analyst (mean $\pm$ SD over runs).}
    \label{tab:alternative-policies}
    \scriptsize
    \setlength{\tabcolsep}{3pt}
    \begin{tabular}{llccc}
        \toprule
        \textbf{\analyst} & \textbf{\judge} & \textbf{Conservative} & \textbf{Relaxed} & \textbf{Excluded} \\
        \midrule
        \gemma & None & 0.583{\tiny\,$\pm$\,0.024} & 0.611{\tiny\,$\pm$\,0.037} & 0.610{\tiny\,$\pm$\,0.029} \\
        \rowcolor{gray!12}  & Self & 0.597{\tiny\,$\pm$\,0.014} & 0.611{\tiny\,$\pm$\,0.024} & 0.610{\tiny\,$\pm$\,0.023} \\
         & Luna & 0.579{\tiny\,$\pm$\,0.029} & 0.606{\tiny\,$\pm$\,0.035} & 0.605{\tiny\,$\pm$\,0.033} \\
        \rowcolor{gray!12}  & Panel$^{\dagger}$ & 0.653{\tiny\,$\pm$\,0.000} & 0.671{\tiny\,$\pm$\,0.016} & 0.671{\tiny\,$\pm$\,0.016} \\
        \midrule
        \qwen & None & 0.667{\tiny\,$\pm$\,0.024} & 0.713{\tiny\,$\pm$\,0.045} & 0.719{\tiny\,$\pm$\,0.035} \\
        \rowcolor{gray!12}  & Self & 0.708{\tiny\,$\pm$\,0.028} & 0.713{\tiny\,$\pm$\,0.058} & 0.725{\tiny\,$\pm$\,0.049} \\
         & Luna & 0.782{\tiny\,$\pm$\,0.029} & 0.801{\tiny\,$\pm$\,0.069} & 0.807{\tiny\,$\pm$\,0.046} \\
        \rowcolor{gray!12}  & Panel$^{\dagger}$ & \textbf{0.819}{\tiny\,$\pm$\,0.064} & \textbf{0.852}{\tiny\,$\pm$\,0.079} & \textbf{0.854}{\tiny\,$\pm$\,0.070} \\
        \midrule
        \luna & None & 0.597{\tiny\,$\pm$\,0.028} & 0.694{\tiny\,$\pm$\,0.028} & 0.800{\tiny\,$\pm$\,0.008} \\
        \rowcolor{gray!12}  & Self & 0.579{\tiny\,$\pm$\,0.080} & 0.704{\tiny\,$\pm$\,0.016} & 0.801{\tiny\,$\pm$\,0.044} \\
         & Haiku & 0.611{\tiny\,$\pm$\,0.000} & 0.708{\tiny\,$\pm$\,0.000} & 0.796{\tiny\,$\pm$\,0.013} \\
        \midrule
        \rowcolor{gray!12} \terra & None & 0.592{\tiny\,$\pm$\,0.043} & 0.690{\tiny\,$\pm$\,0.032} & 0.751{\tiny\,$\pm$\,0.036} \\
         & Self & 0.562{\tiny\,$\pm$\,0.054} & 0.671{\tiny\,$\pm$\,0.046} & 0.726{\tiny\,$\pm$\,0.052} \\
        \rowcolor{gray!12}  & Haiku & 0.617{\tiny\,$\pm$\,0.027} & 0.741{\tiny\,$\pm$\,0.032} & 0.802{\tiny\,$\pm$\,0.027} \\
        \midrule
        \sol & None & 0.576{\tiny\,$\pm$\,0.058} & 0.617{\tiny\,$\pm$\,0.039} & 0.687{\tiny\,$\pm$\,0.028} \\
        \rowcolor{gray!12}  & Self & 0.601{\tiny\,$\pm$\,0.045} & 0.681{\tiny\,$\pm$\,0.045} & 0.772{\tiny\,$\pm$\,0.079} \\
         & Haiku & 0.557{\tiny\,$\pm$\,0.022} & 0.722{\tiny\,$\pm$\,0.012} & 0.808{\tiny\,$\pm$\,0.007} \\
        \midrule
        \rowcolor{gray!12} Claude \haiku & None & 0.579{\tiny\,$\pm$\,0.056} & 0.745{\tiny\,$\pm$\,0.021} & 0.805{\tiny\,$\pm$\,0.034} \\
         & Self & 0.602{\tiny\,$\pm$\,0.032} & 0.759{\tiny\,$\pm$\,0.016} & 0.844{\tiny\,$\pm$\,0.014} \\
        \rowcolor{gray!12}  & Luna & 0.611{\tiny\,$\pm$\,0.024} & 0.718{\tiny\,$\pm$\,0.008} & 0.797{\tiny\,$\pm$\,0.034} \\
        \midrule
        Claude \sonnet & None & 0.593{\tiny\,$\pm$\,0.021} & 0.722{\tiny\,$\pm$\,0.000} & 0.767{\tiny\,$\pm$\,0.000} \\
        \rowcolor{gray!12}  & Self & 0.754{\tiny\,$\pm$\,0.050} & 0.718{\tiny\,$\pm$\,0.008} & 0.781{\tiny\,$\pm$\,0.023} \\
         & Luna & 0.562{\tiny\,$\pm$\,0.018} & 0.741{\tiny\,$\pm$\,0.016} & 0.789{\tiny\,$\pm$\,0.019} \\
        \midrule
        \rowcolor{gray!12} Claude \opus & None & 0.556{\tiny\,$\pm$\,0.050} & 0.718{\tiny\,$\pm$\,0.008} & 0.718{\tiny\,$\pm$\,0.036} \\
         & Self & 0.630{\tiny\,$\pm$\,0.018} & 0.696{\tiny\,$\pm$\,0.047} & 0.700{\tiny\,$\pm$\,0.054} \\
        \rowcolor{gray!12}  & Luna & 0.612{\tiny\,$\pm$\,0.074} & 0.738{\tiny\,$\pm$\,0.001} & 0.775{\tiny\,$\pm$\,0.026} \\
        \bottomrule
    \end{tabular}
    \par\smallskip
    \noindent\begin{minipage}{\columnwidth}\footnotesize\raggedright
        \emph{Self}: the \analyst model acts as \judge. $^{\dagger}$Panel: \qwen, \gemma, \luna, and \muse.
    \end{minipage}
\end{table}

\raggedbottom
\section{Verdict Breakdown}
\label{app:verdict-breakdown}

Figures \ref{fig:verdict-flow-gemma4}--\ref{fig:verdict-flow-claude-opus-5-5}
visualize the verdict of the \analyst in comparison to the different
configurations of the \judge as in \autoref{tab:results}. 
The abbreviations stand for A: affected, P: potentially affected, N: not
affected, and C: cyber-guardrail. Blue represents \dg{1}, orange \dg{2},
green \dg{3}, and gray the cases canceled by a cyber-guardrail without a report. 
The figures are calculated by using a \emph{conservative} policy
(Appendix~\ref{app:policies}).

\input{verdict_flows.tikz.tex}
\ScarvaVerdictFigures

\end{document}